\documentclass{svproc}
\usepackage{graphicx}%
\usepackage{multirow}%
\usepackage{amsmath,amssymb,amsfonts}%
\usepackage{mathrsfs}%
\usepackage[title]{appendix}%
\usepackage{xcolor}%
\usepackage{textcomp}%
\usepackage{manyfoot}%
\usepackage{booktabs}%
\usepackage{algorithm}%
\usepackage{algorithmicx}%
\usepackage{algpseudocode}%
\usepackage{listings}%
\usepackage{url}

\usepackage{float,epsfig}
\usepackage{graphicx}
\usepackage{rotating}
\usepackage{subfigure}
\usepackage{caption}
\usepackage{subcaption}
\usepackage{tkz-graph}
\usepackage{color}
\usepackage{algorithm}
\usepackage{multicol}
\usepackage[utf8]{inputenc}
\usepackage{braket}
\usepackage{subcaption, multirow}
\usepackage[overload]{empheq}
\usepackage{hyperref}
\usepackage{xcolor}
\usepackage{color,graphicx,float,lscape,enumerate}
\usepackage{algorithm}
\def \R{{\mathbb R}}
\def \ep{\epsilon}

\begin{document}
\mainmatter              % start of a contribution
\title{Signed network models for portfolio optimization}
\titlerunning{Signed networks for portfolio optimization}  % abbreviated title (for running head)
\author{Bibhas Adhikari \inst{}} %\and Author2Name Author2Surname\inst{2} Author3Name Author3Surname \and Author4Name Author4Surname \and Author5Name Author5Surname \and Author6Name Author6Surname}
\authorrunning{Bibhas Adhikari} % abbreviated author list (for running head)

\institute{Fujitsu Research of America, Inc.\\
\email{badhikari@fujitsu.com}\\ %WWW home page:
%\texttt{https://sites.google.com/view/bibhas-adhikari/}\\
%\and
%Institution2,
Santa Clara, CA, USA}

\maketitle              % typeset the title of the contribution

\begin{abstract}
In this work, we consider weighted signed network representations of financial markets derived from raw or denoised correlation matrices, and examine how negative edges can be exploited to reduce portfolio risk. We then propose a discrete optimization scheme that reduces the asset selection problem to a desired size by building a time series of signed networks based on asset returns. To benchmark our approach, we consider two standard allocation strategies: Markowitz’s mean–variance optimization and the $1/N$ equally weighted portfolio. Both methods are applied on the reduced universe as well as on the full universe, using two datasets: (i) the Market Champions dataset, consisting of 21 major S\&P500 companies over the 2020–2024 period, and (ii) a dataset of 199 assets comprising all S\&P500 constituents with stock prices available and aligned with Google’s data. Empirical results show that portfolios constructed via our signed network selection perform as good as those from the classical Markowitz model and the equal-weight benchmark in most occasions. 

%and performs better in several occa 

%In this paper, we consider weighted signed network models for representation of financial markets from the correlation or denoised correlation matrix and investigate the potential of negative edges for reducing portfolio risk.  Further, we propose a discrete optimization method for reducing dimension of the portfolio formation problem for finding a potential subset of assets in a financial market for investment by defining a time-series of signed networks from the return values of the assets. Then we employ existing efficient methods such as Markowitz's and $1/N$ to investigate the performance of these methods for optimized portfolio on this low dimensional problem against the performance of these methods for the entire market by considering a market of S\&P500 index of 21 major companies for the time period 2020 to 2024. We observe that our proposed method performs comparatively better than the original Markowitz mean-variance model and $1/N$ method. 

%The abstract should summarize the contents of the paperusing at least 70 and at most 150 words. It will be set in 9-pointfont size and be inset 1.0 cm from the right and left margins.There will be two blank lines before and after the Abstract. \dots
% We would like to encourage you to list your keywords within
% the abstract section using the \keywords{...} command.
\end{abstract}
\section{Introduction}\label{sec1}
%
%The Introduction section, of referenced text \cite{bib1} expands on the background of the work . The introduction should not include subheadings.

%Springer Nature does not impose a strict layout as standard. 

The framework of combinatorial graphs or networks serves as a powerful mathematical tool across a variety of data analysis techniques. In financial applications, networks play a central role in modeling dependencies among assets via their correlation strengths. By representing assets as vertices and encoding correlations as (weighted) edges, numerous methods have been developed for tasks such as asset‐price prediction and risk analysis \cite{chi2010network} \cite{mantegna1999hierarchical} \cite{barigozzi2019nets} \cite{chen2020correlation} \cite{hautsch2015financial} \cite{peralta2016network} \cite{gregnanin2024signature}. Compared to purely statistical approaches, network analysis offers the advantage of capturing both pairwise interactions and higher‐order group dynamics among assets. Several surveys and monographs explore the role of networks in finance and economics more broadly \cite{kenett2015network} \cite{jackson2014networks}  \cite{acemoglu2015systemic}. 

%\cite{nagurney2021networks} 
%\cite{elliott2014financial}
%In these works, a “network” is formalized as a graph $G=(V,E)$, where $V$ is the vertex set and $E\subseteq V\times V$ the edge set. Unless otherwise specified, we restrict attention to simple graphs, i.e.\ graphs with no loops, so that $(v,v)\notin E$ for any $v\in V$.

\begin{figure}[htbp]
    \centering
    \subfigure[]{\includegraphics[width=0.70\textwidth]{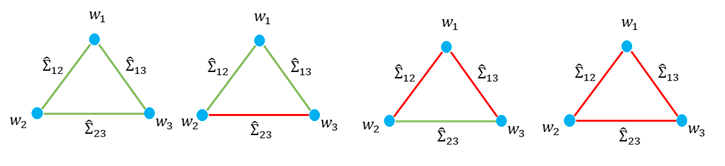}}
    ~ ~ ~
    \subfigure[]{\includegraphics[width=0.20\textwidth]{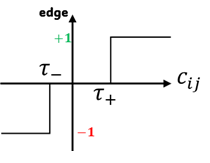}} 
    \caption{(a) Triangles in a signed graph, $T_0,T_1,T_2,T_3$ (from left to right) the green edges are positive and red edges are negative (b) Threshold function \cite{harary2002signed} for signed network formation. $c_{ij}$ denotes the covariance or correlation strength for the assets $i$ and $j$}
    \label{fig:sH_threshold}
\end{figure}

On the other hand, a signed graph augments a standard graph by assigning each edge a sign - positive or negative. When vertices represent random variables, a positive (resp.\ negative) edge indicates that the corresponding variables are positively (resp.\ negatively) correlated. For a comprehensive review of signed‐graph theory and its applications, see Zaslavsky’s annotated bibliography of recent developments \cite{zaslavsky2012mathematical}. Structural balance theory, which hinges on the sign‐configuration of triangles, is fundamental in the study of signed social networks \cite{cartwright1956structural} \cite{harary1953notion}. Triangles are classified by the number of negative edges they contain: if $T_j$ denotes a triangle with $j$ negative edges for $j=0,1,2,3$, then $T_0$ and $T_2$ are balanced, whereas $T_1$ and $T_3$ are unbalanced (see Figure \ref{fig:sH_threshold}(a)). A signed graph is called \emph{balanced} if its vertex set can be partitioned into two subsets such that every positive edge lies within a subset and every negative edge connects vertices across subsets \cite{harary1953notion}. Empirical evidence shows that real‐world signed networks are typically unbalanced, inspiring various measures to quantify this lack of balance \cite{aref2019balance} \cite{singh2017measuring} \cite{diaz2025mathematical}. 

Harary et al. \cite{harary2002signed} introduced the notion of balance signed graphs for well-structured equities portfolios that could contain risk in the portfolio. In their model, assets are considered as vertices, and the existence of positive and negative edges in the corresponding signed graph is defined by the correlation between returns of the associated pair of assets. Thus the edges indicate the tendency or manner in which the value of the assets change relative to each other. A positive edge between a pair of assets reflects that the valuation of the assets tend to move in tandem, whereas a negative edge implies that the valuations of the assets move in opposite direction, if one goes up the other goes down. Following the idea of Harary et al., a number of articles considered to investigate financial markets through signed graph models and vice versa, for instance see \cite{huffner2010separator}  \cite{aref2019balance} \cite{ehsani2020structure} \cite{figueiredo2014maximum}  \cite{vasanthi2015applications} and the references therein.    Recently, 
in \cite{bartesaghi2025global}, the authors show that the global balance index of financial correlation networks can be
used as a systemic risk measure. We note that, even though weighted correlation networks are considered in several context in the literature, weighted signed network models for financial networks are rare to find \cite{masuda2025introduction}.

%Incorporating denoising methods for numerical correlation strengths into Harary et al.'s approach, we propose modeling financial networks as weighted signed graphs. Further, we propose normalizing the edge weights using a threshold function to define a signed network model, which we term normalized market or portfolio graphs. Since the edges in this signed graph represent tendency of positively and negatively correlated assets, we argue the existence of unbalance triangles in such normalized graphs. Indeed, we propose to determine the values of the parameters in the denoising method and threshold function in such a way that the number of unbalance triangles is minimized in the normalized market or portfolio graph. We further discuss the weighted signed graph models in the context of Markowitz's portfolio theory, and we show that triangles with at least one negative edges reduces portfolio risk more than triangles with all positive edges. 

Despite its elegance, Markowitz’s portfolio construction is plagued in practice by estimation errors in the covariance matrix and expected return. Consequently, an optimized portfolio based on an estimated covariance matrix will almost surely deviate from the true Markowitz solution \cite{goldberg2022dispersion}. To reduce the dimension of the problem, we propose a discrete optimization problem by incorporating co-movement statistics of asset returns over all times $t\in\{1,\hdots, T\}$, thus generalizing Kendall’s Tau \cite{kendall1938new}. Our two‐step framework for designing a diversified, hedge‐protected portfolio is as follows:

\begin{enumerate}
    \item {\bf Dimensionality reduction via signed‐graph models.}
We construct a time series of signed graphs on the asset set by comparing each pair’s returns to their own sample means over a rolling window of length $T$. An edge is assigned a positive sign if both returns lie on the same side of their means, and a negative sign otherwise. Negative edges therefore capture hedge relationships, instances where returns move in opposite directions, directly targeting variance reduction without explicit covariance estimation. We then score assets by the frequency with which they exhibit negative edges against others, and select the top candidates by maximizing these weighted counts.

\item {\bf Final allocation on the reduced universe.}
Having selected a smaller asset subset, we apply any standard allocation method such as Markowitz’s mean–variance model and the $1/N$ rule \cite{demiguel2009optimal} to compute the investment weights.
\end{enumerate}

By filtering the investment universe in the first step, we  reduce problem size while retaining hedge-relevant information, and then leverage established optimization techniques on this subset with a desired number of assets. To demonstrate our findings on real financial data, we consider two datasets. The first is the Market Champions dataset from \href{https://www.kaggle.com/datasets/jijagallery/industry-leaders-performance-dataset}{Kaggle}
, which contains daily stock prices for 21 prominent S\&P 500 companies across multiple sectors, covering the period from January 1, 2020 to December 31, 2024. The second dataset is from \href{https://www.kaggle.com/datasets/paultimothymooney/stock-market-data/data}{Kaggle}
 and consists of 199 S\&P 500 stocks with price data aligned with Google’s dataset. We evaluate the performance of our proposed method through backtesting and observe that it performs comparably to both the standard Markowitz optimization and the $1/N$ equally weighted strategy applied to the full set of assets in the financial market.
 
Finally, note that quantum computing constitutes a fundamentally novel paradigm for portfolio optimization. A spectrum of quantum algorithmic frameworks including quantum annealing, variational quantum algorithms such as the Quantum Approximate Optimization Algorithm (QAOA) has been employed to address Markowitz’s mean–variance problem \cite{hegade2022portfolio} \cite{leipold2024train} \cite{kordonowy2025lie} \cite{soloviev2025scaling}. These approaches are intrinsically designed for large‐scale instances, however, their implementation on fault‐tolerant hardware remains a future prospect. In contrast, Noisy Intermediate‐Scale Quantum (NISQ) platforms enable empirical assessment of both purely quantum and hybrid quantum-classical algorithms on moderately sized portfolios \cite{buonaiuto2023best}. By integrating our dimension‐reduction methodology with the operational capacity of NISQ devices, we show a promising pathway for the practical realization of hybrid quantum-classical portfolio optimization.

%Finally, we mention that quantum computing is a transcendental technique for solving portfolio optimization problems. Several quantum algorithmic paradigms such as quantum annealers, Variational Quantum Algorithms such as Quantum Approximate Optimization algorithms (QAOA), and quantum walks based algorithms are proposed to solve Markowitz's mean-variance models \cite{hegade2022portfolio} \cite{IMBQ}. These methods are specially designed to deal with large scale portfolio optimization problems, though  the unavailability of fault-tolerant quantum computers restrict these methods to be executed in quantum hardwares. Indeed,  Noisy Intermediate Scale Quantum (NISQ) devices allows testing the performances of both quantum and hybrid algorithms of moderate scale problems for portfolio optimization \cite{buonaiuto2023best}. Our proposed dimension reduction technique along with availability of NISQ devices offers a new avenue of performing hybrid algorithms for execution of quantum algorithms for portfolio optimization.  

%The rest of the paper is organized as follows. In Section \ref{sec2}, we discuss the importance of negative edges in a portfolio to contain risk in the formulation of signed network representation of a financial market. In Section \ref{sec3}, we propose a new dimension reduction method for portfolio optimization. The empirical analysis is performed in Section \ref{sec4}.
%based on hedge statistics of assets in a market through a time-series of signed network models. 

\section{Financial markets as signed graphs}\label{sec2}

%In this section we consider weighted signed graph models for representing financial markets using correlation matrices.  %We analytically show that negative edges in these graphs reduce the portfolio risk due to Markowitz's mean-variance model for portfolio optimization \cite{markowitz1952portfolio}.   

%\subsection{Weighted signed networks as market graphs}

%Harary et al. introduced the notion of balance signed graph into the formation of portfolio graphs of investors for risk management. They are the first to emphasize that the formation of well-structured portfolio need not to include assets with only positively correlated returns of the assets. Their suggestion was to consider the return of the assets relative to each other within the portfolio, especially include assets in a portfolio that are negatively correlated with some assets which work as hedges in the portfolio. 

In this section, we consider weighted signed graph models for representing financial markets using correlation matrices.  
Since the actual correlation between the returns is unobserved, the correlation is often estimated by employing several statistical estimators \cite{mantegna1999introduction}. Denoting the unobserved covariance matrix as $\Sigma$ for a random vector $\boldsymbol{R}=(R_1,\hdots,R_N),$ we denote an estimator of $\Sigma$ as $\widehat{\Sigma}=[\widehat{\Sigma}_{ij}],$ where $\widehat{\Sigma}_{ij}=\mbox{Cov}(R_i,R_j)=E[(R_i-\mu_{R_i})(R_j-\mu_{R_j})]$ denotes the estimated covariance corresponding to the random variables $R_i$ and $R_j.$ Here, $\mu_{X}=E[R]$, the expected value of the random variable $R.$ In financial time-series data, let $R_i(t)$ denote the random variable corresponding to an index associated with the asset $i$ at time $t$ (for example, a day or month or year). Then a popular unbiased estimator for $\Sigma$ is the sample covariance matrix, whose entries are defined by $\widehat{\Sigma}_{ij}=\frac{1}{T-1}\sum_{t=1}^T (r_i^t-\mu_{R_i})(r_j^t-\mu_{R_j}),$ where $R_i(t)=r_i^t$ and $R_j(t)=r_j^t,$ $\mu_R=\frac{1}{T}\sum_{t=1}^T r^t,$ and $t\in\{1,\hdots, T\}$ with $T$ is the total time window. The sample correlation coefficient matrix is then defined as $\widehat{\rho}=[\widehat{\rho}_{ij}],$  with $\rho_{ij}=\mbox{Cov}(R_i,R_j)/\sqrt{\mbox{Var}(R_i)\mbox{Var}(R_j)},$ where $\mbox{Var}(X)=\frac{1}{T-1}\sum_{t=1}^T (x^t-\mu_X)^2$ is nonzero, and $\widehat{\rho}$ estimates the  population Pearson correlation matrix. Note that $-1\leq \widehat{\rho}_{ij}\leq 1$ with $\widehat{\rho}_{ij}=1$ if $i=j.$ If $\widehat{\rho}_{ij}>0$ then the random variables $X_i$ and $X_j$ are said to be positively correlated and they are negatively correlated if $\widehat{\rho}_{ij}<0.$

For financial time-series data, such as in stock market, let $S_n(t)$ denote the random variable for the price of the $n$-th stock at time $t.$ Then the random variable $R_n(t)$ which represents return of the $n$-th stock for a fixed time horizon $\Delta t$ is defined as: 
$(S_n(t+\Delta t)-S_n(t))/S_n(t)$ (Linear return) or $\log S_n(t+\Delta t) - \log S_n(t)$ (log return). Often the value of $\Delta t$ is considered as $1$.  For Markowitz's portfolio theory applications, a correlation coefficient estimator matrix must be non-singular, and hence positive definite. We mention here that there are other powerful methods to model the return time-series, such as the GARCH process introduced by Bollerslev \cite{bollerslev1986generalized}, a generalization of the ARCH process proposed by Engle in \cite{engle1982autoregressive}.

%Then the Pearson correlation matrix for returns of the stocks for a time period $T$ is defined as $\widehat{\Sigma}(T)=\frac{1}{T}MM^\dagger,$ where $M=[M_{kt}]\in\mathbb{R}^{K\times T}$ with $M_{kt}=\left(R_k(t)-E[{R_k]}\right)/\sqrt{E[R_k^2] - E[R_k]^2},$ where $E[{R_k}]=\frac{1}{T}\sum_{t=1}^T R_k(t)$ and $E[R_k^2]=\frac{1}{T}\sum_{t=1}^T R_k(t)^2.$ Thus, by definition, $\widehat{\Sigma}(T)$ is a positive semi-definite matrix. 

However, it is demonstrated in literature that for finite time-series data i.e. when $T<\infty,$ there is a random offset to every correlation coefficient and these values are dressed up with noise \cite{guhr2003new}, it can be validated by comparing eigenvalue density of a correlation matrix to a random matrix \cite{laloux1999noise}. An important observation from the financial data is that the effect of noise strongly depends on the ratio $N/T$, where $N$ is the size of the portfolio and $T$ the length of the available time series \cite{pafka2003noisy}, see also \cite{kondor2007noise}\cite{chung2022effects}. 

% \cite{pafka2004estimated} 

%There are several noise reduction methods proposed in the literature such as the power mapping method .

%Let $\widehat{\Sigma}(T)=[\widehat{\Sigma}_{ij}]$ be a estimated noisy correlation matrix corresponding to a financial market. The the noise-reduced correlation matrix $\widehat{\Sigma}^{(q)}=[\widehat{\Sigma}_{ij}^{(q)}]$ corresponding to the power mapping method is defined as $\widehat{\Sigma}_{ij}^{(q)}=\mbox{sign}(\widehat{\Sigma}_{ij})\left|\widehat{\Sigma}_{ij}\right|^{q},$ where $q>1.$ Since the diagonal entries of $\widehat{\Sigma}(T)$ are $1,$ they are not affected by the powering map. Obviously, the magnitude of the off-diagonal correlations are reduced and the reduction depends on the value of $q.$ It is also demonstrated that a suitable value of the parameter $q$ depends on the time horizon $T$ and the power method is robust in a sense that it yields good results for a wide range of values of $q$ such as $q\in\{1.25. 1.50, 1.75\}$ \cite{schafer2010power}. 

%In this paper, we propose a criteria to choose the value of $q$ based on the normalized signed graph representation of the denoised correlation matrix defined in squeal. 

The weighted signed graph $G^s({\widehat{\Sigma}_D}),$ which represents a model financial market associated with a (denoised) correlation estimator matrix $\widehat{\Sigma}_D=[\widehat{\Sigma}_{ij}^{D}]$, is defined as follows. 

\begin{definition}\label{def:wsg}(Weighted signed graph models of financial markets)
The vertex set of $G^s({\widehat{\Sigma}_D})$ is the set of assets in a portfolio index by $1,2,\hdots, N.$ Then the edge set $E\subseteq V\times V$ is defined by the two following ways.    
\begin{enumerate}
    \item Without thresholding: there is an edge between a pair of vertices $(i,j)$ if and only if $\widehat{\Sigma}_{ij}^D\neq 0.$ The sign of an edge $(i,j)$ is positive if $\widehat{\Sigma}_{ij}^D> 0$ and negative if $\widehat{\Sigma}_{ij}^D< 0.$ The weight of the edge is $\widehat{\Sigma}_{ij}^D.$ 
    \item With thresholding: let $0<\tau_+<1$ and $-1<\tau_+<0.$ Then There is a positive edge for the vertex pair $(i,j)$ with weight $\widehat{\Sigma}_{ij}^D$ if $\widehat{\Sigma}_{ij}^D>\tau_+$ and a negative edge for the vertex pair $(i,j)$ with weight $\widehat{\Sigma}_{ij}^D$ if $\widehat{\Sigma}_{ij}^D<\tau_-.$
\end{enumerate}
\end{definition}

%The challenges associated with denoising and thresholding techniques are to decide the values of the concerned parameters: the level of denoising parameters, the threshold parameters. % $\tau_+,$ $\tau_-.$ We discuss an optimization method to select the values of these parameters in the next section. 

%\subsection{Signed networks as normalized market graphs}

A signed graph representation of a financial market is the underlying signed graph obtained by relaxing the edge weights of a weighted signed portfolio graph. This can be achieved in two ways: directly from the estimated correlation matrix with thresholding and from the denoised correlation matrix. In both cases, the threshold function may or may not be applied.  In \cite{harary2002signed}, Harary et al. considered using a threshold function directly from the estimated correlation matrix as described in Figure \ref{fig:sH_threshold} (b). As they explained, the edges in the normalized market graph represent the tendency of the return values of the associated assets (vertices).  

%Thus, the resulting normalized signed graph from a correlation matrix or denoised correlation matrix whose edges represent the tendency of return values of the incident pair of assets  to go `up' or `down' in tandem for positive edges and negative edges to imply go in opposite directions i.e. return values of one goes `up' and the other goes `down'. In what follows, we argue that the the normalized signed graphs as defined by Hararey et al. should not have a triangle of type $T_1$ and $T_3.$ In fact, we argue that this observation incites a fundamental difference between normalized social networks and financial networks. 

%The thresholding technique is explained in Fig. \ref{fig:sH_threshold} (b).

In social signed network systems, structural balance theory plays a pivotal role to investigate the dynamics of the underlying systems and it is believed that social networks evolve toward balance, however it may not be true in all real-world social networks \cite{diaz2025mathematical}. It is also demonstrated using real-world data that the number of unbalance triangles $T_1$ and $T_3$ is significantly lesser than the number of balance triangles $T_0$ and $T_2$. In financial normalized networks, if the edges represent tendencies of going up or down of the return values, then for a triangle of type $T_3$ of three assets $X,Y,Z$ would mean the following: if the return of $X$ goes up then the returns of $Y$ of $Z$ must go down (due to the negative edges $(X,Y)$ and $(X,Z)$), however if both the return values of $Y$ and $Z$ go down then they must be positively correlated which contradicts the fact that they are negatively correlated. Thus, the crucial point here is the rates of going up or down of the pairs of return values, which are decided by the correlation values. A similar argument can also be given for the existence of $T_1$ type triangles. Thus we conclude that structural properties of financial (unweighted) signed networks is strikingly different from social signed networks.

Now we establish from the viewpoint of containing risk that negative edges in a signed graph representation plays an important role to contain portfolio risk than a portfolio with all positive edges (positively correlated assets). We consider the variance of the portfolio as a measure of risk from the perspective of Markowitz's portfolio theory (MPT) \cite{markowitz1952portfolio} \cite{mattera2023shrinkage}. According to MPT, for a diversified portfolio, an investor's goal is to minimize the portfolio variance where the minimimum-variance portfolio problem can be written as $\min_w w^\dagger\widehat{\Sigma} w$ such that ${\bf 1}^\dagger w=1,$ %\begin{equation}\label{eqn:min_var_port}
%\begin{cases}
 %   \min_w w^\dagger\widehat{\Sigma} w \\
  %  s.t. \,\, {\bf 1}^Tw=1,
%\end{cases}    
%\end{equation} 
where $\widehat{\Sigma}$ is the risky assets' (return) estimated covariance matrix, $w=[w_1, \hdots, w_N]^T,$ $w_j\geq 0$ is the vector of portfolio weights i.e.  the proportion of wealth invested in the assets, and ${\bf 1}$ is the all-one vector of dimension $N,$ the number of total number of assets in the portfolio. The condition $w_j\geq 0$ means that the portfolio does not contain any short positions. Then we have the following theorem.

%Next we show that for any portfolio weight vector  $w=[w_1,\hdots,w_N]\in R^N_{\geq 0}$ for a portfolio of $K$ assets with a covariance estimator $\widehat{\Sigma}=[\widehat{\Sigma}_{ij}]\in\mathbb{R}^{N\times N}$, the risk volatility $w^T\widehat{\Sigma}w$ is lesser when $\widehat{\Sigma}_{ij}\in \mathbb{R}$ and the underlying normalized signed graph contains at least one negative edge than the risk volatility corresponding to the covariance matrix $|\widehat{\Sigma}|,$ whose entries are $|\widehat{\Sigma}_{ij}|\in\mathbb{R}_{\geq 0}.$

\begin{theorem}\label{thm:var}
Let $w=[w_1,\hdots,w_N]^\dagger$ with $w_i\geq 0$ and $\sum_{i=1}^N w_i=1.$ Suppose $G^s(\widehat{\Sigma})$ is the underlying (weighted) signed graph with at least one negative edge. Then $w^\dagger \widehat{\Sigma}w\leq w^\dagger |\widehat{\Sigma}|w,$ where  $|\widehat{\Sigma}|=[|\widehat{\Sigma}_{ij}|]$
\end{theorem}

\begin{proof} The proof follows from the fact that $$w^\dagger \widehat{\Sigma}w=\sum_{i=1}^N 
w_i^2\widehat{\Sigma}_{ii} + \sum_{\substack{i\neq j\\ i,j=1}}^N2\, \mbox{sign} (\widehat{\Sigma}_{ij})\, |\widehat{\Sigma}_{ij}| w_iw_j.$$

%    Let us denote the $(i,j)$-th entry of $\widehat{\Sigma}$ is denoted as $\widehat{\Sigma}_{ij}=\mbox{sgn}(\widehat{\Sigma}_{ij})\, |\widehat{\Sigma}_{ij}|,$ where $\mbox{sgn}(x)$ denotes the sign of $x.$ Note that by definition, $$w^\dagger \widehat{\Sigma}w =\sum_{i=1}^K w_i^2\widehat{\Sigma}_{ii} + \sum_{\substack{i\neq j\\ i,j=1}}^K2\, \mbox{sign}(\widehat{\Sigma}_{ij})\, |\widehat{\Sigma}_{ij}| w_iw_j.$$ Then the desired result follows from the fact that $\mbox{sign}(\widehat{\Sigma}_{ij})<0$ for at least one pair $(i,j),$ whereas it is positive for $|\widehat{\Sigma}|.$ \hfill{$\square$}
\end{proof} 

Triangles constitute a fundamental motif, prevalent both in social networks and in correlation‐based financial networks. In the context of portfolio construction, a natural question then arises: which triangle configurations in the weighted signed graph of asset returns most effectively contribute to variance reduction and thus help contain risk? From Figure \ref{fig:sH_threshold} (a), it is easy to verify that $w^\dagger \widehat{\Sigma}w$ is minimum for the (unbalance) triangle $T_3$ when $w_i>0$ for all $i.$ Indeed, when short selling is allowed, it becomes a different story. In the case of short selling, the portfolio weights corresponding to assets which hold short positions are considered negative. Then in Figure \ref{fig:sH_threshold} (a), considering $w_1<0$ and $w_2, w_3$ to be positive it follows that the unbalance triangle $T_1$ achieves the minimum portfolio risk. However, changing the assignment of signs of the edges but keeping the balance/unbalance property of the triangles fixed, the minimum risk could be achieved by a different type of triangle.

%Structural balance theory for signed graphs is grounded in the classification of triangles as either balanced or unbalanced.

%Harary et al. argued that triangles of types $T_2$ can induce hedging in the portfolio and hence diversify the portfolio more than type $T_0$ triangles, and it protects the portfolio against market shocks in the form of adverse changes in the value of the assets. Now, we briefly analyze the effect of weighted signed triangles to contain portfolio risk. 

%of three assets without short-selling or in a portfolio with triangles, a crucial question is which type of triangles contribute to reduce the portfolio risk more than others.

\section{Signed network based hedge-protected  portfolio formation}\label{sec3}

Theorem \ref{thm:var} affirms that negative edges act like hedges in a portfolio, as defined in \cite{baur2010gold}. Now note that the sample covariance of return values of a pair of assets is given by $\widehat{\Sigma}_{ij}=\frac{1}{T-1}\sum_{t=1}^T (R_i^t-\mu_{R_i})(R_j^t-\mu_{R_j})$ for a time period $T$, where $R_k^t$ denotes the return of asset $k$ at time $t,$ and $\mu_{R_k}$ is the mean of the return values of the asset $k$ for the time period $T.$ If $\widehat{\Sigma}_{ij}<0,$ it indicates that one of the assets had a few `bad days' compare to its own mean return value than the other asset in terms of their  return values, although for the other days their return values could be at per compare to their own mean return values. Whereas, if $\widehat{\Sigma}_{ij}>0$ then it would mean that they have the same `bad days' and `good days' i.e. return values of both the assets go up or down together corresponding to their own mean return values in most of the days or the values go up or down quite deep together on a few days compare to the days when pairwise go in opposite directions making a pair (up,down) or (down,up). In an extreme case, one ``very good" or ``very bad" day of either or both the assets can flip the sign of $\widehat{\Sigma}_{ij}$ from positive to negative or vice-versa. By compressing these finer co‐movement patterns into $\widehat{\Sigma}$, the Markowitz mean-variance formulation masks this local return dynamics. This interpretation applies equally to raw and denoised (or thresholded) covariance estimators; henceforth, ``covariance matrix" refers to either form.

Recall that the original mean-variance model (OMV) model is formulated as 
\begin{eqnarray}
    \mbox{OMV1:} && w^* = \arg\min_{w\in\Delta_K} w^\dagger \widehat{\Sigma}w \,\, \mbox{s.t.} \mu^\dagger w=\ep \label{eqn:OMV1}\\
    \mbox{or OMV2:} && w^* = \arg\min_{w\in\Delta_K} -\mu^\dagger w + \gamma w^\dagger\widehat{\Sigma}w, \label{eqn:OMV2}
\end{eqnarray} where $\mu$ is the mean vector consists of the means of the asset returns and $\Delta_K=\left\{w\in\R^N_{\geq 0} : \sum_{i=1}^N w_i =1\right\}$ \cite{lai2022survey}. Thus Markowitz's model recommends formation of portfolio to ensure some level of $\ep$ (also called target return) of portfolio return $\mu^\dagger w$ and minimizing the portfolio variance given by equation (\ref{eqn:OMV1}), and simultaneously maximizing the return and minimizing the portfolio variance simultaneously with a mixing parameter (also called \textit{risk aversion parameter}) $\gamma\in (0,\, \infty)$ in equation (\ref{eqn:OMV2}). Thus, both the models urge to gain more return and withstand less risk. From the computational perspective, note that this is a convex optimization problem and efficient methods are available to solve such optimization problems. Indeed, considering $w\in \R^N$ (with short selling), the analytical solution of problem OMV2 is given by \cite{boyd2024markowitz} 
\begin{equation}\label{eqn:MPsol}
w^*=\frac{1}{2\gamma}\widehat{\Sigma}^{-1}(\mu+\nu^*{\bf 1}), \,\, \nu^*=\frac{2\gamma-{\bf 1}^\dagger\widehat{\Sigma}^{-1}\mu}{{\bf 1}^\dagger \widehat{\Sigma}^{-1}{\bf 1}}.    
\end{equation}

In the weighted graph representation of a portfolio, we observe that a negative edge helps to reduce portfolio risk. As proved in Theorem \ref{thm:var}, for any invest allocation vector, the risk can be contained more by having negative edges (negatively correlated assets) than positively correlated edges (positively correlated assets) of equal strengths. Observing this, we define \textit{hedge score} of an asset in a portfolio by introducing a time-series of portfolio graphs for a time period $T$ and the negative degrees of the vertices as follows.

We define a normalized market signed graph $G^s_t(\boldsymbol{\mu,R_N})=(V, E_t)$ of $N$ assets at a time $t\in\{1,\hdots,T\}$ with $V=\{1,\hdots,N\}$ as the set of assets, $\mu$ is the mean return vector of the assets and $\boldsymbol{R_N}=(R_1^t,\hdots,R_N^t)$ is the observed empirical return values. The edge set is defined as follows. For a pair of assets $(i,j)$ there is a positive edge if $(R_i^t-\mu_{R_i})(R_j^t-\mu_{R_j})\geq 0$ and a negative edge if $(R_i^t-\mu_{R_i})(R_j^t-\mu_{R_j})< 0.$ Then based on the statistics of negative degree (the number of negative edges at a vertex is adjacent to) of an asset, we have the following definition preserving Markowitz's model through the proposed time-varying normalized market graph representation.

\begin{definition}\label{def:hs}(Hedge score)
Let $S^t_{n}:V\setminus\{n\}\rightarrow \{0,1\}$ be a function $S^t_n(j)=1$ if $(R_n^t-\mu_{R_n})(R_j^t-\mu_{R_j})<0$ and $S^t_n(j)=0$ otherwise, where $t\in\{1,\hdots,T\}$. Then the hedge score of an asset $n$ is defined as
\begin{equation}\label{eqn:hs}
H(n,T)= \frac{\sum_{\substack{j\in V\\ j\neq k}} \sum_{t=1}^{T} S_n^t(j)}{T(N-1)}. 
\end{equation}
\end{definition}

Note that $S_n^t$ counts the negative degree of the vertex $n$ in the graph $G^s_t(\mu,R_N).$ Besides, $0\leq H(n,T)\leq 1.$ Then we propose the following optimization problem for selecting a potential subset of assets for the design of a hedge-protected diversified portfolio as follows. 
\begin{equation}\label{eqn:opts}
    \mbox{OPT:} \,\, \arg\max_{S\subseteq V} \sum_{n\in S} H(n,T)\mu_{R_n}(T)=\arg\max_{S\subseteq V} H_S^\dagger(T) \mu_S(T),
\end{equation} where the mean value of the returns of an asset $i$ for a time period $T$ and $N$ is the total number of assets in the market. For a set of assets $S\subseteq V,$ denote the $H_S(T)=[H(s_1,T), \hdots, H(s_{|S|},T)]^\dagger$ and $\mu(S,T)==[\mu_{R_{s_1}}(T), \hdots, \mu_{R_{s_{|S|}}}(T)]^\dagger$ as the column vectors of hedge scores and mean return values respectively, of the assets $s_k\in S$ within a time period $T.$ Note the theoretical maximum value of equation (\ref{eqn:opts}) would be given by the complete graph $G^s_t(\boldsymbol{\mu,R_N})$ will all edges are negative for all $t,$ however, from financial data such a graph can never be realized for moderate size value of $|S|.$ 

We mention that the time complexity of solving the optimization problem is $O(N\log N),$ which follows from the fact that the indices $H(k,T)\mu_k(T)$ can be stored in an array of length $N$ and the optimizer $S$ is then obtained by sorting this array. We could add a  constraint $|S|=K\leq N$ to equation (\ref{eqn:opts}) and   the value of $K$ could be decided by the investor's input and by performing portfolio risk analysis of the potential choices. Overall, this optimization method significantly reduces the dimension of investment allocation problem. Once the set $S$ of assets is determined by solving the optimization problem, the invest allocation vector $w$ can be chosen by employing methods such as $1/|S|$ method or the original Markowitz's mean-variance method.  

\begin{figure}[htbp]
    \centering
    {\includegraphics[width=0.90\textwidth]{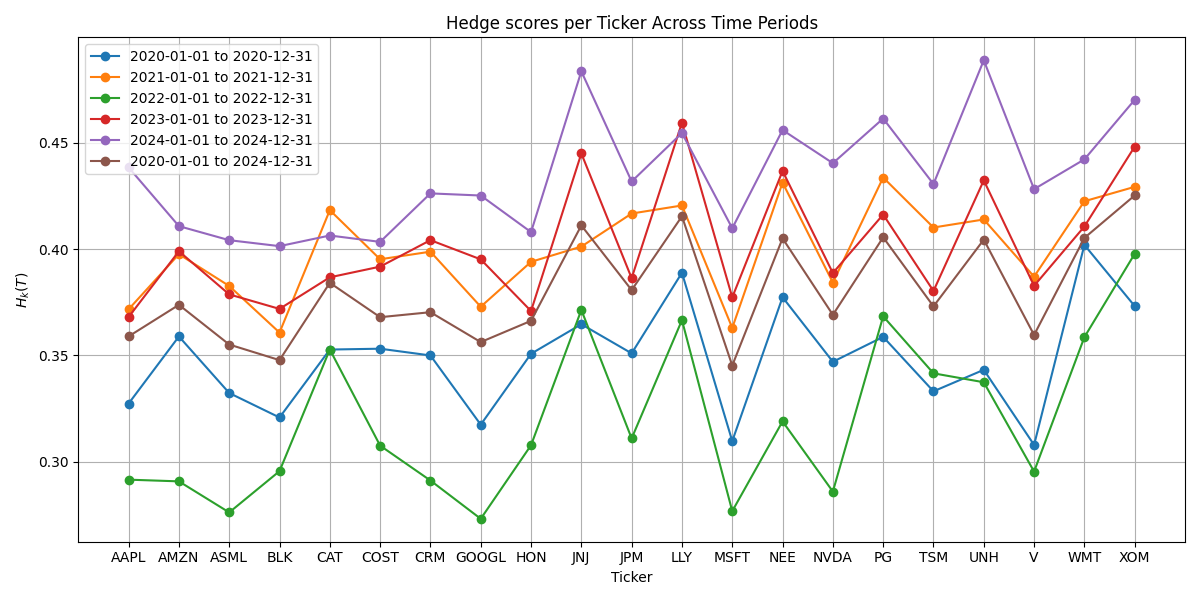}} 
    %\subfigure[]{\includegraphics[width=0.24\textwidth]{monalisa.jpg}}
    %\subfigure[]{\includegraphics[width=0.24\textwidth]{monalisa.jpg}}
    \caption{Hedge scores of all tickers during 2020 to 2024}
    \label{fig:sggeneration}
\end{figure}

\section{Empirical analysis}\label{sec4}

To test the proposed methodology for portfolio construction, we consider two datasets. First we consider a data of S\&P500 index from January 1, 2020, to December 31, 2024, available online in \href{https://www.kaggle.com/datasets/jijagallery/industry-leaders-performance-dataset}{Kaggle}. This data contains stock prices of major companies, called Market Champions: Leading Stocks Dataset, from different sectors including Technology \& AI: Apple (AAPL), Microsoft (MSFT), Alphabet (GOOGL), Amazon (AMZN), NVIDIA (NVDA), Taiwan Semiconductor (TSM), Healthcare: Johnson \& Johnson (JNJ), UnitedHealth Group (UNH), Eli Lilly (LLY), Energy: ExxonMobil (XOM), NextEra Energy (NEE), Financial: JPMorgan Chase (JPM), Visa (V), BlackRock (BLK). Consumer: Walmart (WMT), Costco (COST), Procter \& Gamble (PG), Industrial: Caterpillar (CAT), Honeywell (HON). Software/Cloud: Salesforce (CRM), ASML Holding (ASML). 

%\bigskip
%\begin{verbatim}
\begin{table}[]
    \centering
    \caption{The stock selection based on the proposed optimization method for the Market Champions dataset, setting $K\in\{5,8,12,15\}$ The set $\mathcal{S}_K$ of assets for a year 20XX is obtained by using the data of all the 252 days in the year 20XX. }
    \begin{tabular}{|c|c|c|c|c|}
    \toprule
      Year   & $\mathcal{S}_5$ & $\mathcal{S}_8$ & $\mathcal{S}_{12}$ & $\mathcal{S}_{15}$ \\ \midrule \hline
       2020  & \tiny{AAPL, AMZN, ASML, NVDA, TSM} & $\mathcal{S}_5$ with &  $\mathcal{S}_8$ and &  $\mathcal{S}_{12}$ and \\
         &   & \tiny{BLK, CRM, MSFT} &  \tiny{COST, GOOGL, LLY, NEE} & \tiny{CAT, UNH, WMT} \\ \hline \hline 
        2021  & \tiny{ASML, GOOGL, LLY, NVDA, XOM}  & $\mathcal{S}_5$ and & $\mathcal{S}_8$ and & $\mathcal{S}_{12}$ and \\
        &  & \tiny{COST, MSFT, UNH} & \tiny{AAPL, BLK, JPM, NEE} & \tiny{CAT, CRM, PG} \\ \hline\hline
        2022 & \tiny{CAT, HON, LLY, UNH, XOM} & $\mathcal{S}_5$ and & $\mathcal{S}_8$ with & $\mathcal{S}_{12}$ and \\
        & &\tiny{JNJ, V, WMT} & \tiny{COST, JPM, NEE, PG} & \tiny{AAPL, BLK, CAT}\\ \hline \hline
        2023 & \tiny{AMZN, CRM, GOOGL, LLY, NVDA} & $\mathcal{S}_5$ and & $\mathcal{S}_8$ and & $\mathcal{S}_{12}$ and \\
        &&\tiny{AAPL, COST, MSFT} & \tiny{ASML, JPM, TSM, V} & \tiny{BLK, CAT, WMT} \\ \hline\hline
        2024 & \tiny{AMZN, JPM, NVDA, TSM, WMT} & $\mathcal{S}_5$ and & $\mathcal{S}_8$ with & $\mathcal{S}_{12}$ and \\
        && \tiny{AAPL, COST, GOOGL} & \tiny{BLK, CAT, CRM, LLY} & \tiny{MSFT, NEE, V}\\\hline%\bottomrule
    \end{tabular}
    \label{tab:my_label}
\end{table}
%\bigskip
%\end{verbatim}

Forming the signed graphs $G^s_t(\boldsymbol{\mu,R_N})$, for each day $t$ the hedge-score for each asset is calculated following equation (\ref{eqn:hs}) for the time period $T$, which is considered a year such as 2020, 2021, 2022, 2023, 2024, and for the entire period January $2020$ to December $2024$ in Figure \ref{fig:sggeneration}. Based on these hedge score statistics of all the assets and setting $|S|=K\in\{5,8,12,15\},$ we determine the potential subset $\mathcal{S}_K$ of 21 assets in the market by solving equation (\ref{eqn:opts}) for each time period $T,$ described in Table \ref{tab:my_label}. 

Next we consider the dataset of the entire collection of assets whose data are aligned with the Google stock in S\&P500 index from August 2004 to Dec 2022. This data forms a universe of 199 assets, available in \href{https://www.kaggle.com/datasets/paultimothymooney/stock-market-data/data}{Kaggle}. The assets are given by 'A', 'AAP', 'ABMD', 'ABT', 'ACN', 'ADI', 'ADM', 'ADP', 'ADSK', 'AJG', 'AKAM', 'ALB', 'ALGN', 'ALK', 'AMAT', 'AMD', 'AME', 'AMGN', 'AMT', 'AMZN', 'AOS', 'APA', 'APD', 'ARE', 'ATVI', 'AVY', 'BAC', 'BAX', 'BBY', 'BDX', 'BEN', 'BIIB', 'BIO', 'BRK-A', 'BSX', 'BWA', 'BXP', 'CAG', 'CB', 'CCI', 'CDE', 'CHD', 'CHRW', 'CINF', 'CLX', 'CMI', 'CNC', 'COO', 'COP', 'CPB', 'CPRT', 'CRM', 'CSCO', 'CTAS', 'CTSH', 'CUK', 'D', 'DGX', 'DOV', 'DPZ', 'DVA', 'EA', 'EBAY', 'ECL', 'EFX', 'EL', 'EMN', 'ES', 'EW', 'EXR', 'FAST', 'FIS', 'FISV', 'FITB', 'FLS', 'FMC', 'FTI', 'GGG', 'GILD', 'GIS', 'GOOG', 'GPC', 'GPN', 'GWW', 'HAS', 'HBAN', 'HD', 'HES', 'HRB', 'HRL', 'HST', 'HSY', 'HUM', 'IDXX', 'IFF', 'ILMN', 'ISRG', 'ITW', 'IVZ', 'JBHT', 'JCI', 'JKHY', 'JNPR', 'JPM', 'K', 'KIM', 'KMB', 'KSS', 'LEG', 'LH', 'LNC', 'LNT', 'LOW', 'MAA', 'MAR', 'MCHP', 'MCO', 'MDLZ', 'MLM', 'MMC', 'MOS', 'MSFT', 'NEE', 'NEOG', 'NFLX', 'NI', 'NOC', 'NOV', 'NTAP', 'NTRS', 'NVR', 'NWL', 'O', 'ODFL', 'OMC', 'ORLY', 'OXY', 'PAYX', 'PCAR', 'PH', 'PHM', 'PKG', 'PKI', 'PLD', 'PNW', 'PPG', 'PRU', 'PVH', 'RCL', 'REG', 'RF', 'RHI', 'RLI', 'ROK', 'ROL', 'ROP', 'SBUX', 'SCHW', 'SEE', 'SHW', 'SIVB', 'SLB', 'SLG', 'SNPS', 'SO', 'SPG', 'SRE', 'STT', 'SWK', 'SYK', 'T', 'TJX', 'TMO', 'TRV', 'TSCO', 'TSN', 'TTWO', 'TXT', 'TYL', 'UDR', 'URI', 'VFC', 'VMC', 'VRSN', 'VZ', 'WAT', 'WBA', 'WDC', 'WEC', 'WHR', 'WM', 'WMB', 'WRB', 'WST', 'WYNN', 'XEL', 'YUM', 'ZBH', 'ZION'. As in the preceding dataset, we determine potential asset sets $\mathcal{S}_n$ for $n=20$ and $50$ based on solving the optimization problem stated in equation (\ref{eqn:opts}). These scores are computed from the signed graphs $G^s_t(\boldsymbol{\mu},\boldsymbol{R}_N)$ constructed for each day $t$ over a time period $T$, where $T$ corresponds to one year for each of the years from 2005 to 2022. We report the reduced universe obtained using the proposed optimization scheme in in Table \ref{tab:sp5001}.

\begin{table}[h!]
    \centering
     \caption{Comparison of backtesting results for the stock market data. Proposed Method (PM), Markowitz's Portfolio with short selling (MP), Markowitz's Portfolio with no short selling (MPNS), Equally Weighted Portfolio (EWP). } 
    \begin{tabular}{|c|c|c|c|c|c|c|c|c|c|}
   \toprule
        Year & Method &  \multicolumn{2}{|c|}{Total Return }  &  \multicolumn{2}{|c|}{Annual Return }  & \multicolumn{2}{|c|}{Annual Volatility }  & \multicolumn{2}{|c|}{Sharpe Ratio} \\
        && $K=5$ & $K=8$ & $K=5$& $K=8$ & $K=5$ & $K=8$ & $K=5$ & $K=8$ \\ \midrule  \hline %\hline
        %\multicolumn{3}{|c|}{Header for all columns} 
        2021 & PM+MP&-50& 298.38&-61&153.56&39.94&53.99&-1.54& 2.84 \\
        & PM+MPNS&102.97&102.97&81.09&81.09&44.77&44.77&1.81&1.81 \\
        & PM+EWP& 35.81&33.94&34.31&31.87&26.72&22.47&1.28 &1.42\\\cline{2-10}
        & MP &\multicolumn{2}{|c|}{-7.98}&\multicolumn{2}{|c|}{-7.25}&\multicolumn{2}{|c|}{14.80}&\multicolumn{2}{|c|}{-0.49} \\
        & MPNS & \multicolumn{2}{|c|}{102.97}&\multicolumn{2}{|c|}{81.09}&\multicolumn{2}{|c|}{44.77}&\multicolumn{2}{|c|}{1.81} \\
        & EWP &\multicolumn{2}{|c|}{29}&\multicolumn{2}{|c|}{26.55}&\multicolumn{2}{|c|}{13.91}&\multicolumn{2}{|c|}{1.91} \\ \hline\hline
        2022 & PM+MP &-58.59& 14.39&-74.77&17.19&53&27.02&-1.41& 0.64 \\
         & PM+MPNS &-60.26&-60.26&-72.90&-72.90&63.23&63.23&-1.15&-1.15  \\
         & PM+EWP &-18.49&-17.90&-15.10&-15.69&33.23&28.98&-0.45&-0.54 \\\cline{2-10}
         & MP &\multicolumn{2}{|c|}{-15.05}&\multicolumn{2}{|c|}{-14.71}&\multicolumn{2}{|c|}{18.57}&\multicolumn{2}{|c|}{-0.79} \\
         &MPNS &\multicolumn{2}{|c|}{-60.26}&\multicolumn{2}{|c|}{-72.90}&\multicolumn{2}{|c|}{63.23}&\multicolumn{2}{|c|}{-1.15} \\
         & EWP &\multicolumn{2}{|c|}{-19.76}&\multicolumn{2}{|c|}{-19.05}&\multicolumn{2}{|c|}{25.07}&\multicolumn{2}{|c|}{-0.76} \\\hline\hline
        2023 & PM+MP &-17.52&-7.15&-17.13&-6.77&21.73&12.16&-0.79&-0.56 \\
         & PM+MPNS &-8.97&-8.97&-6.40&-6.40&24.99&24.99&-0.26&-0.26 \\
        & PM+EWP &11.84&9.82&12.38&10.17&14.56&11.81&0.85&0.86 \\\cline{2-10}
        & MP & \multicolumn{2}{|c|}{-14.46}& \multicolumn{2}{|c|}{-15.04}& \multicolumn{2}{|c|}{12.42}& \multicolumn{2}{|c|}{-1.21} \\
        &MPNS & \multicolumn{2}{|c|}{-8.97}& \multicolumn{2}{|c|}{-6.40}& \multicolumn{2}{|c|}{24.99}& \multicolumn{2}{|c|}{-0.26} \\
        & EWP &\multicolumn{2}{|c|}{ 29.81}& \multicolumn{2}{|c|}{27.27}& \multicolumn{2}{|c|}{13.09}& \multicolumn{2}{|c|}{ 2.08}\\ \hline\hline
        2024 & PM+MP & -100&-26.39& -434.13&-22.38& 368.12&41.09& -1.18&-0.54 \\
        & PM+MPNS & 149.38 &149.38&105.72 &105.72& 52.10&52.10&2.03 &2.03 \\
        & PM+EWP &  53.91&44.42&46.40&38.97&24.07&19.43& 1.93&2.01 \\\cline{2-10}
        & MP &\multicolumn{2}{|c|}{23.60}& \multicolumn{2}{|c|}{21.77} & \multicolumn{2}{|c|}{8.99}& \multicolumn{2}{|c|}{ 2.42} \\
        & MPNS &\multicolumn{2}{|c|}{149.38}& \multicolumn{2}{|c|}{105.72} & \multicolumn{2}{|c|}{52.10}& \multicolumn{2}{|c|}{2.03} \\
        & EWP &\multicolumn{2}{|c|}{28.84} & \multicolumn{2}{|c|}{26.31}& \multicolumn{2}{|c|}{12.35}& \multicolumn{2}{|c|}{2.13}\\ \bottomrule
    \end{tabular}
    %Total Return $=\frac{\mbox{Final Value} \,-\, \mbox{Initial Value}}{\mbox{Initial Value}}\times 100$, Annual Return = $\left[\left(\frac{\mbox{Final Value}}{\mbox{Initial Value}}\right)^{1/n}-1\right]\times 100,$ where $n$ is the number of years, }
    \label{tab:comparison}
\end{table}

\begin{table}[h!]
    \centering
    \caption{The stock selection based on the proposed optimization method, setting $K\in\{20,50\}$ from the S\&P500 dataset of 199 stocks. The set $\mathcal{S}_K$ of assets for a year 20XX is obtained by using the data of all the 252 days in the year 20XX. }
    \begin{tabular}{c|cl|}
    \toprule
    \hline
        Year & \multicolumn{2}{|c|}{reduced universe} \\ \midrule \hline 
         2005 & $\mathcal{S}_{20}:$ & \tiny{'ISRG', 'NFLX', 'GOOG', 'CRM', 'NOV', 'HUM', 'WDC', 'CCI', 'HES', 'GPN', 'AKAM',  } \\
         && \tiny{'ILMN', 'SLB', 'AMT','GILD', 'WMB', 'WRB', 'AAP', 'OXY', 'FLS'} \\
         & $\mathcal{S}_{50}:$ & $\mathcal{S}_{20}$ with 
          \tiny{'APA', 'MLM', 'FTI', 'AMD', 'MCO', 'COP', 'TSCO', 'DPZ', 'ORLY', 'A', } \\
         && \tiny{'PRU', 'BEN', 'EFX', 'IDXX', 'CHRW', 'RHI', 'ROP', 'DVA', 'FAST', 'SLG',}\\
         && \tiny{ 'VMC', 'ATVI', 'CB', 'SCHW', 'IVZ', 'AMGN', 'PHM', 'SRE','MCHP', 'URI'}  \\ \hline \hline
         2006 & $\mathcal{S}_{20}:$ & \tiny{'ILMN', 'AKAM', 'ALGN', 'WST', 'SLG', 'ALB', 'WYNN', 'TYL', 'CSCO', 'PVH','IVZ', }\\
         && \tiny{'ABMD', 'VFC', 'BXP', 'TMO', 'CTSH', 'KSS', 'T', 'IFF', 'MOS'} \\
         & $\mathcal{S}_{50}:$ & $\mathcal{S}_{20}$ with \tiny{'ES', 'NTAP', 'MAR', 'FMC', 'LH', 'SHW', 'AMT', 'FTI', 'HAS', 'CAG', } \\
         && \tiny{ 'PCAR', 'LNT', 'ADM', 'KIM', 'CPB','NEE', 'UDR', 'MLM', 'MDLZ', 'CPRT', 'SNPS', } \\ %\hline\hline
         && \tiny{'VMC', 'CMI', 'WAT', 'SCHW', 'SPG', 'REG', 'VZ', 'HST', 'ACN'}\\ \hline\hline 
         2007 & $\mathcal{S}_{20}:$ & \tiny{'MOS', 'ISRG', 'NOV', 'AMZN', 'HES', 'CMI', 'NEOG', 'FTI', 'FLS', 'ATVI', 'CRM', }\\
         && \tiny{'JNPR', 'SLB', 'APA',  'WAT', 'OXY', 'IDXX', 'VRSN', 'BWA', 'ILMN' }\\
         & $\mathcal{S}_{50}:$ & $\mathcal{S}_{20}$ with \tiny{'WDC', 'GOOG', 'TXT', 'PH', 'ADM', 'GILD', 'AME', 'FMC', 'CPRT',}\\
         && \tiny{ 'WMB', 'HUM', 'BRK-A', 'APD', 'SYK', 'IVZ', 'CCI', 'COP', 'YUM', 'SCHW', 'TMO', } \\ %\hline\hline
         && \tiny{'MLM','ALGN', 'BIO', 'ROL', 'BAX', 'PCAR', 'PKG', 'CHD', 'NEE', 'CHRW'} \\ \hline\hline
         2008 & $\mathcal{S}_{20}:$ &\tiny{'AMGN', 'ODFL', 'EW', 'ALK', 'HRB', 'HAS', 'NFLX', 'GILD', 'RLI', 'ABMD', }\\
         && \tiny{ 'AJG', 'GIS', 'WRB', 'CHRW','TSCO','SHW',  'CHD', 'PHM', 'WM', 'JBHT' }\\
          & $\mathcal{S}_{50}:$ & $\mathcal{S}_{20}$ with \tiny{'DGX', 'LOW', 'SO', 'TYL', 'ROL', 'ABT', 'ORLY', 'WST', 'BAX', }\\
          && \tiny{'GWW', 'O', 'ADP', 'NEOG', 'VMC', 'ACN','LEG', 'MMC', 'FAST', 'AAP', 'HSY', }\\ %\hline\hline
          && \tiny{'HD', 'AOS', 'NVR', 'DVA', 'MAA', 'CLX', 'CB', 'ILMN', 'WEC', 'TRV'}\\ \hline\hline
         2009 & $\mathcal{S}_{20}:$ & \tiny{'AMD', 'WDC', 'AMZN', 'NTAP', 'SBUX', 'CTSH', 'ISRG', 'CRM', 'FTI', 'CCI',}\\
         && \tiny{ 'COO', 'ALGN', 'CDE', 'NFLX', 'SLG', 'PVH', 'GOOG', 'A', 'WHR', 'MOS' }\\
         & $\mathcal{S}_{50}:$ & $\mathcal{S}_{20}$ with \tiny{'DPZ', 'TJX', 'TYL', 'EMN', 'WAT', 'FLS', 'EW', 'RCL', 'NOV', 'AKAM',}\\
         && \tiny{  'ADI', 'GPN', 'NVR', 'SIVB','PKG', 'EBAY', 'IVZ', 'PRU', 'MSFT', 'JCI', }\\ %\hline\hline
         && \tiny{'BEN', 'CMI', 'HST', 'WBA', 'ALB', 'SPG', 'APD', 'IDXX', 'MCHP', 'KSS'}\\\hline \hline 
         2010 & $\mathcal{S}_{20}:$ & \tiny{'NFLX', 'URI', 'ILMN', 'CMI', 'BWA', 'EW', 'DPZ', 'AKAM', 'ZION', 'HBAN', }\\
         && \tiny{ 'TSCO', 'CRM', 'RCL', 'NEOG', 'AAP','ODFL', 'EL', 'ALK', 'NTAP', 'WYNN'} \\
         & $\mathcal{S}_{50}:$ & $\mathcal{S}_{20}$ with \tiny{'ORLY', 'COO', 'PH', 'CTSH', 'CDE', 'PVH', 'PCAR', 'ROL', 'AME', }\\
         && \tiny{ 'FTI', 'FITB', 'ADSK', 'TSN', 'HAS', 'NOV','HST', 'MAR', 'ROK', 'EXR', 'UDR', }\\ %\hline\hline
         && \tiny{'ALB', 'ROP', 'FAST', 'HRL', 'AMZN', 'FMC', 'SBUX', 'YUM', 'GWW', 'SHW'} \\ \hline\hline
          2011 & $\mathcal{S}_{20}:$ & \tiny{ 'DPZ', 'ABMD', 'BIIB', 'ISRG', 'HUM', 'TJX', 'VFC', 'CNC', 'TSCO', 'EL',}\\
         && \tiny{  'TYL', 'SBUX', 'FAST', 'HSY', 'RLI','ORLY', 'NI', 'CHD', 'HRB', 'WMB'}\\
         & $\mathcal{S}_{50}:$ & $\mathcal{S}_{20}$ with \tiny{'ALK', 'EXR', 'GWW', 'URI', 'EA', 'TSN', 'COO', 'CPRT', 'D', 'SPG', }\\
         && \tiny{ 'SO', 'MCO', 'ODFL', 'WRB', 'YUM','CTAS', 'ABT', 'MDLZ', 'ALGN', 'HD', 'FTI',}\\ %\hline\hline
         && \tiny{ 'CAG', 'KMB', 'WEC', 'LNT', 'AMGN', 'AMT', 'XEL', 'NEE', 'GIS'}\\\hline \hline
         2012 & $\mathcal{S}_{20}:$ & \tiny{'PHM', 'WHR', 'BAC', 'ILMN', 'GILD', 'SHW', 'CRM', 'EBAY', 'CCI', 'TYL', 'EMN', }\\
         && \tiny{'RF', 'URI', 'PVH', 'PPG', 'PKG', 'EXR', 'AOS', 'PKI', 'HD'}\\
         & $\mathcal{S}_{50}:$ & $\mathcal{S}_{20}$ with \tiny{'DVA', 'NEOG', 'WST', 'LOW', 'AMZN', 'MCO', 'FLS', 'WDC', 'EFX',  }\\
         && \tiny{'NWL', 'AMGN', 'TJX', 'AMT', 'NFLX','NVR', 'JBHT', 'TMO', 'COO', 'AME', 'FIS',}\\ %\hline\hline
         && \tiny{ 'FISV', 'TXT', 'SRE', 'DPZ', 'TSCO', 'FMC', 'BAX', 'RCL', 'BIIB', 'VMC'}\\ \hline \hline
         2013 & $\mathcal{S}_{20}:$ & \tiny{'NFLX', 'BBY', 'ABMD', 'BSX', 'ILMN', 'GILD', 'ALGN', 'TYL', 'BIIB','WDC',} \\
         && \tiny{  'SEE', 'SIVB', 'WST', 'LNC','TSN', 'TSCO', 'SCHW', 'ALK', 'VFC', 'ATVI' }\\
         & $\mathcal{S}_{50}:$ & $\mathcal{S}_{20}$ with \tiny{'WYNN', 'TMO', 'AOS', 'NOC', 'EA', 'URI', 'PKG', 'AMD', 'HUM', 'PRU', }\\
         && \tiny{'GOOG', 'TTWO', 'WBA', 'ADM', 'AAP', 'HRB', 'HES', 'AMZN', 'HAS', 'NEOG', }\\ %\hline\hline
         && \tiny{'TJX', 'FIS', 'DPZ', 'AMAT', 'FLS', 'VRSN', 'CNC', 'ODFL', 'MCO', 'BWA'}\\\hline 
    \end{tabular}
   % \caption{Collection of hedge-protected assets for S\&P500 data of a total 199 assets}
    \label{tab:sp5001}
\end{table}

\begin{table}[ht]
    \centering
    \begin{tabular}{c|cl|}
    \hline
      %  Year & \multicolumn{2}{|c|}{Sets of hedge-protected assets} \\ \hline 
      2014 & $\mathcal{S}_{20}:$ &  \tiny{'EW', 'EA', 'CNC', 'RCL', 'ILMN', 'ALK', 'TTWO', 'MAR', 'ODFL', 'ORLY',} \\
         && \tiny{ 'ABMD', 'ARE', 'AAP', 'ISRG', 'SHW','IDXX', 'AMAT', 'EXR', 'REG', 'HUM'}\\
         & $\mathcal{S}_{50}:$ & $\mathcal{S}_{20}$ with \tiny{'LOW', 'RHI', 'AMGN', 'WBA', 'WEC', 'PNW', 'XEL', 'DPZ', 'LEG', 'UDR', }\\
         && \tiny{'LNT', 'WDC', 'ES', 'NI', 'AKAM', 'O', 'COO', 'CHRW', 'URI', 'BXP', }\\ %\hline\hline
         && \tiny{'DGX', 'SLG', 'NVR', 'NEE', 'CTAS', 'NOC', 'SPG', 'SRE', 'APD', 'KIM'}\\ \hline \hline
         2015 & $\mathcal{S}_{20}:$ & \tiny{'ABMD', 'NFLX', 'AMZN', 'ATVI', 'TYL', 'GPN', 'EXR', 'EA', 'HRL', }\\
         && \tiny{ 'GOOG','SBUX', 'VRSN', 'VMC', 'BSX', 'TSN','ALK', 'EFX', 'AOS', 'NVR', 'CRM'} \\
         & $\mathcal{S}_{50}:$ & $\mathcal{S}_{20}$ with \tiny{'ORLY', 'CNC', 'RLI', 'HUM', 'CUK', 'HAS', 'TTWO', 'NOC', 'FISV', 'JKHY',}\\
         && \tiny{ 'EW', 'JNPR', 'HD', 'MLM', 'RCL', 'UDR', 'CLX', 'MDLZ', 'AVY', 'PKI',  }\\ %\hline \hline
         && \tiny{'ROL', 'MAA', 'CPB', 'ALGN', 'DPZ', 'ROP', 'NI', 'NEOG', 'CAG', 'MSFT'}\\ \hline \hline 
         2016 & $\mathcal{S}_{20}:$ & \tiny{'CDE', 'AMD', 'AMAT', 'IDXX', 'MLM', 'ZION', 'ALB', 'CMI', 'RF', 'DPZ', }\\
         && \tiny{'SIVB', 'URI', 'ALGN', 'ODFL', 'TTWO', 'WST', 'FMC', 'APA', 'CPRT', 'BBY'}\\
         & $\mathcal{S}_{50}:$ & $\mathcal{S}_{20}$ with \tiny{'PH', 'MCHP', 'WM', 'FITB', 'JBHT', 'NTAP', 'PKG', 'BAC', 'ABMD', 'PCAR', }\\
         && \tiny{'VMC', 'SYK', 'BIO', 'JPM', 'LNC', 'COO', 'ROL', 'ITW', 'HES', 'ROK', } \\ %\hline
         && \tiny{'ADI', 'WYNN', 'DGX', 'PRU', 'CINF', 'CTAS', 'AKAM', 'SNPS', 'ADM', 'AJG'}\\ \hline\hline
       2017 & $\mathcal{S}_{20}:$ & \tiny{'ALGN', 'TTWO', 'NVR', 'WYNN', 'PHM', 'CNC', 'ATVI', 'ILMN', 'ISRG', 'ABMD', }\\
         && \tiny{ 'FMC', 'EL', 'BBY', 'MAR', 'AVY','AMAT', 'AMZN', 'NTAP', 'GGG', 'PVH'} \\
         & $\mathcal{S}_{50}:$ & $\mathcal{S}_{20}$ with
         \tiny{'URI', 'MCO', 'CPRT', 'NFLX', 'ODFL', 'VRSN', 'BAX', 'SHW', 'ABT', 'CRM',}\\
         && \tiny{'SWK', 'AME', 'ALB', 'GPN', 'SNPS', 'A', 'RCL', 'HD', 'WAT', 'VFC',} \\ %\hline\hline
         && \tiny{'PKG', 'ROL', 'ROK', 'PH', 'PKI', 'ROP', 'MCHP', 'AMT', 'ADSK', 'SIVB'}\\ \hline\hline 
         2018 & $\mathcal{S}_{20}:$ &  \tiny{'AMD', 'ABMD', 'AAP', 'ORLY', 'CHD', 'DPZ', 'BSX', 'NFLX', 'EW', 'VRSN',}\\
         && \tiny{'ILMN', 'CRM', 'AMZN', 'ISRG', 'GWW', 'ABT', 'KSS', 'ADSK', 'HRL','AJG',}\\
         & $\mathcal{S}_{50}:$ & $\mathcal{S}_{20}$ with
           \tiny{'TJX', 'IDXX', 'MSFT', 'RLI', 'ROL', 'COO', 'NEE', 'TMO', 'HUM', 'COP', }\\
         && \tiny{'AMT', 'ADP', 'O', 'CNC', 'YUM', 'FISV', 'SBUX', 'TSCO', 'CSCO','AMGN', 'CPRT', }\\ %\hline\hline
         &&\tiny{'MOS', 'ECL', 'JKHY', 'FIS', 'CLX', 'NTAP', 'CTAS', 'WEC', 'EXR'} \\\hline \hline
         2019 & $\mathcal{S}_{20}:$ &  \tiny{'AMD', 'CDE', 'CPRT', 'AMAT', 'TSN', 'GPN', 'MLM', 'NVR', 'WDC', 'TYL', }\\ 
         && \tiny{'CAG', 'BBY', 'MCO', 'CPB', 'FISV', 'HES', 'BIO', 'SNPS', 'EL', 'PLD' }\\
         & $\mathcal{S}_{50}:$ & $\mathcal{S}_{20}$ with
          \tiny{'DOV', 'AMT', 'EW', 'CTAS', 'ODFL', 'SO', 'PHM', 'FMC', 'WST', 'NEE', }\\
         &&\tiny{'URI', 'VMC', 'MSFT', 'HSY', 'MAA', 'DVA', 'APD', 'SHW', 'SRE', 'ACN', }\\ %\hline\hline
         &&\tiny{'ZBH', 'EFX', 'AKAM', 'WEC', 'GIS', 'VFC', 'IDXX', 'ARE', 'TMO', 'AVY'} \\\hline \hline
         2020 & $\mathcal{S}_{20}:$ &  \tiny{'AMD', 'ABMD', 'WST', 'ALB', 'IDXX', 'TTWO', 'NFLX', 'AMZN', 'SNPS', 'ALGN', }\\
         && \tiny{'ROL', 'ATVI', 'BIO', 'ADSK', 'TSCO', 'DVA', 'ODFL', 'TYL', 'SIVB', 'PKI'} \\
         & $\mathcal{S}_{50}:$ & $\mathcal{S}_{20}$ with
          \tiny{'TMO', 'CLX', 'EBAY', 'EA', 'DPZ', 'AMAT', 'CDE', 'MSFT', 'A', 'CRM', }\\
         && \tiny{ 'GGG', 'ISRG', 'EFX', 'CPRT', 'URI','LOW', 'NEE', 'CHD', 'CTSH', 'FAST', }\\ %\hline\hline 
         &&\tiny{'GOOG', 'SHW', 'AJG', 'MCHP', 'EL', 'ABT', 'CTAS', 'PH', 'CMI', 'EMN'} \\ \hline\hline
         2021 & $\mathcal{S}_{20}:$ &  \tiny{'EXR', 'MAA', 'SPG', 'AMAT', 'APA', 'COP', 'ODFL', 'PLD', 'RHI', 'TSCO', }\\
         && \tiny{ 'GOOG', 'SIVB', 'KIM', 'WST', 'OXY','REG', 'JCI', 'MOS', 'UDR', 'AMD'} \\
         & $\mathcal{S}_{50}:$ & $\mathcal{S}_{20}$ with
          \tiny{'LOW', 'TXT', 'DPZ', 'ACN', 'MLM', 'HD', 'SCHW', 'MSFT', 'JNPR', 'LH',}\\
         && \tiny{  'ALB', 'ORLY', 'FITB', 'EFX', 'AOS','HRB', 'AAP', 'MMC', 'WAT', 'EW', }\\ %\hline\hline
         &&\tiny{'JBHT', 'PAYX', 'DGX', 'BAC', 'TMO', 'NVR', 'SEE', 'SHW', 'WM', 'SNPS'} \\ \hline\hline
         2022 & $\mathcal{S}_{20}:$ &  \tiny{'OXY', 'FTI', 'HES', 'HRB', 'APA', 'COP', 'SLB', 'NOV', 'NOC', 'ADM', }\\
         && \tiny{ 'WRB', 'GIS', 'CPB', 'GPC', 'HSY','WMB', 'SRE', 'AMGN', 'GILD', 'MOS' }\\
         & $\mathcal{S}_{50}:$ & $\mathcal{S}_{20}$ with
          \tiny{'TRV', 'ORLY', 'BIIB', 'K', 'HUM', 'RLI', 'ROL', 'GWW', 'PCAR', 'FMC', }\\
         && \tiny{'AJG', 'CAG', 'PNW', 'CB', 'ATVI', 'ALB', 'CMI', 'BSX', 'URI', 'CTAS', }\\ %\hline\hline
         && \tiny{'JKHY', 'APD', 'ADP', 'CNC', 'XEL', 'ABMD', 'TJX', 'SO', 'OMC', 'WM'}\\ \hline
    \end{tabular}
    %\caption{Year-wise collection of hedge-protected assets for S\&P500 data of a total 199 assets}
    \label{tab:sp5002}
\end{table}

We employ the backtesting method for analyzing the performance of the proposed method and the results are given in Table \ref{tab:comparison} and Table \ref{tab:sp500comparison2}. We determine the optimized portfolio allocation vector solving the Markowitz's with short selling and no short selling optimization problems as described by OMV1 and OMV2 respectively. Using the data of the previous year we form the portfolios and test the performance of these portfolios using standard statistics for its performance for the next year. For instance, we use the stock data of 2020 to form the portfolio and test the performance by finding the Total return (\%), annual return (\%), annual volatility (\%), and the Sharpe ratio using the data of 2021.  Then we employ these methods to construct portfolio from the set of assets $\mathcal{S}_K$ and calculate the above mentioned statistics for these portfolios. We consider the estimators $\widehat{\mu}$ and $\widehat{\Sigma}$ as the sample mean and sample variance of the data. For OMV1, we set the target return ($\epsilon$) as the maximum of the mean return values for the Market Champions data of 21 leading stocks, and average of the 75-quartile mean returns for the S\&P500 data of 199 assets. We also derive the above mentioned statistics for the portfolio allocation vector using the $1/N$ method, we call the associated portfolio as equally weighted portfolio (EWP). We observe that the proposed dimension reduction technique  along with either Markowitz or EWP gives better results compare to employing these methods on all assets of the entire market in several occasions in both the datasets.

\begin{table}[ht]
    \centering
     \caption{Comparison of backtesting results for the stock market data. Proposed Method (PM), Markowitz's Portfolio with short selling (MP), Markowitz's Portfolio with no short selling (MPNS), Equally Weighted Portfolio (EWP). } 
    \begin{tabular}{c|c|c|c|c|c|c|c|c|c|}
    \hline\toprule
        Year & Method &  \multicolumn{2}{|c|}{Total Return}  &  \multicolumn{2}{|c|}{Annual Return}  & \multicolumn{2}{|c|}{Annual Volatility}  & \multicolumn{2}{|c|}{Sharpe Ratio} \\
        && $n=20$ & $n=50$ & $n=20$& $n=50$ & $n=20$ & $n=30$ & $n=20$ & $n=50$ \\ \hline\hline
        %\multicolumn{3}{|c|}{Header for all columns} 
        2006 & PM+MP&6.60& 35.85&8.67&32.70&21.08&18.98&0.41&1.72 \\
        & PM+MPNS&4.27&9.55&5.60&10.23&16.65&14.38&0.34&0.71\\
        & PM+EWP&14.67&10.88&15.30&11.66&17.31&15.83&0.88&0.74\\\cline{2-10}
        & MP &\multicolumn{2}{|c|}{16.70}&\multicolumn{2}{|c|}{16.65}&\multicolumn{2}{|c|}{14.73}&\multicolumn{2}{|c|}{1.13} \\
        & MPNS &\multicolumn{2}{|c|}{11.51}&\multicolumn{2}{|c|}{11.73}&\multicolumn{2}{|c|}{12.18}&\multicolumn{2}{|c|}{0.96} \\
        & EWP&\multicolumn{2}{|c|}{14.11}&\multicolumn{2}{|c|}{14.00}&\multicolumn{2}{|c|}{11.77}&\multicolumn{2}{|c|}{1.19} \\ \hline\hline
        2007 & PM+MP &-24.37&-98.50&-23.48&-177.54&30.45&208.87&-0.77&-0.85 \\
         & PM+MPNS &-14.10&-8.23&-13.34&-7.44&19.88&15.56&-0.67&-0.48  \\
         & PM+EWP &2.28&3.21&4.32&4.84&20.22&18.19&0.21&0.27\\ \cline{2-10}
         & MP &\multicolumn{2}{|c|}{6.21}&\multicolumn{2}{|c|}{7.02}&\multicolumn{2}{|c|}{13.74}&\multicolumn{2}{|c|}{0.51} \\
         &MPNS &\multicolumn{2}{|c|}{-3.46}&\multicolumn{2}{|c|}{-2.56}&\multicolumn{2}{|c|}{14.02}&\multicolumn{2}{|c|}{-0.18} \\
         & EWP &\multicolumn{2}{|c|}{1.89}&\multicolumn{2}{|c|}{3.19}&\multicolumn{2}{|c|}{16.13}&\multicolumn{2}{|c|}{0.20} \\\hline\hline
        2008 & PM+MP &-59.40&-12.45&-0.09&136.89&132.80&173.87&-0.00&0.79 \\
        & PM+MPNS &-42.40&-36.24&-46.43&-39.05&41.50&34.23&-1.12&-1.14 \\
        & PM+EWP &-54.55&-47.83&-63.75&-54.43&54.40&45.74&-1.17&-1.19 \\\cline{2-10}
        & MP &\multicolumn{2}{|c|}{-39.26}&\multicolumn{2}{|c|}{-41.67}&\multicolumn{2}{|c|}{40.48}&\multicolumn{2}{|c|}{-1.03} \\
        &MPNS &\multicolumn{2}{|c|}{-34.84}&\multicolumn{2}{|c|}{-37.20}&\multicolumn{2}{|c|}{33.32}&\multicolumn{2}{|c|}{-1.12} \\
        & EWP &\multicolumn{2}{|c|}{-40.62}&\multicolumn{2}{|c|}{-42.87}&\multicolumn{2}{|c|}{42.72}&\multicolumn{2}{|c|}{-1.00} \\ \hline\hline
        2009 & PM+MP &-5.50&8.57&0.49&10.20&34.42&19.71&0.01&0.52 \\
        & PM+MPNS &6.67&5.45&8.23&6.86&18.65&17.46&0.44&0.39 \\
        & PM+EWP &0.99&6.87&3.86&9.42&23.96&23.43&0.16&0.40 \\\cline{2-10}
        & MP &\multicolumn{2}{|c|}{17.19}&\multicolumn{2}{|c|}{18.63}&\multicolumn{2}{|c|}{23.26}&\multicolumn{2}{|c|}{0.80}  \\
        & MPNS &\multicolumn{2}{|c|}{8.96}&\multicolumn{2}{|c|}{9.92}&\multicolumn{2}{|c|}{16.11}&\multicolumn{2}{|c|}{0.62} \\
        & EWP &\multicolumn{2}{|c|}{20.75}&\multicolumn{2}{|c|}{24.07}&\multicolumn{2}{|c|}{32.03}&\multicolumn{2}{|c|}{0.75} \\ \hline\hline
        2010 & PM+MP &-370.66&105.98&728.86&140.46&385.55&116.49&1.89&1.21 \\
        & PM+MPNS &1.45&22.66&4.74&22.81&25.68&21.40&0.18&1.07 \\
        & PM+EWP &26.25&24.22&26.49&24.43&24.81&22.96&1.07&1.06 \\\cline{2-10}
        & MP &\multicolumn{2}{|c|}{8.70}&\multicolumn{2}{|c|}{9.97}&\multicolumn{2}{|c|}{17.80}&\multicolumn{2}{|c|}{0.56}  \\
        & MPNS &\multicolumn{2}{|c|}{19.94}&\multicolumn{2}{|c|}{19.75}&\multicolumn{2}{|c|}{17.24}&\multicolumn{2}{|c|}{1.15} \\
        & EWP &\multicolumn{2}{|c|}{19.08}&\multicolumn{2}{|c|}{19.44}&\multicolumn{2}{|c|}{19.50}&\multicolumn{2}{|c|}{1.00} \\ \hline\hline
        2011 & PM+MP &-55.83&-99.74&-26.59&-280.29&105.30&241.04&-0.25&-1.16 \\
        & PM+MPNS &-26.76&-10.64&-27.77&-8.92&26.24&21.73&-1.06&-0.41 \\
        & PM+EWP &-17.29&-8.16&-13.96&-4.11&31.81&29.71&0.44&-0.14 \\\cline{2-10}
        & MP &\multicolumn{2}{|c|}{42.73}&\multicolumn{2}{|c|}{37.62}&\multicolumn{2}{|c|}{19.34}&\multicolumn{2}{|c|}{1.94}  \\
        & MPNS &\multicolumn{2}{|c|}{-7.27}&\multicolumn{2}{|c|}{-5.48}&\multicolumn{2}{|c|}{20.38}&\multicolumn{2}{|c|}{-0.27} \\
        & EWP &\multicolumn{2}{|c|}{-4.96}&\multicolumn{2}{|c|}{-1.75}&\multicolumn{2}{|c|}{25.83}&\multicolumn{2}{|c|}{-0.07} \\ \hline\hline 
        2012 & PM+MP &55.10&-0.31&62.36&0.94&60.00&15.52&1.04&0.06 \\
        & PM+MPNS &19.99&5.70&19.41&6.05&13.81&9.39&1.41&0.64 \\
        & PM+EWP &10.72&11.09&11.23&11.35&13.59&11.89&0.83&0.95 \\\cline{2-10} 
        & MP &\multicolumn{2}{|c|}{-2.96}&\multicolumn{2}{|c|}{-0.02}&\multicolumn{2}{|c|}{24.17}&\multicolumn{2}{|c|}{-0.00}  \\
        & MPNS &\multicolumn{2}{|c|}{5.70}&\multicolumn{2}{|c|}{6.05}&\multicolumn{2}{|c|}{9.39}&\multicolumn{2}{|c|}{0.64} \\
        & EWP &\multicolumn{2}{|c|}{14.67}&\multicolumn{2}{|c|}{14.80}&\multicolumn{2}{|c|}{13.76}&\multicolumn{2}{|c|}{1.08} \\ \hline
    \end{tabular}
    %\caption{SP500 comparison, setting target return as average of the 75‐quartile mean returns } 
    %Total Return $=\frac{\mbox{Final Value} \,-\, \mbox{Initial Value}}{\mbox{Initial Value}}\times 100$, Annual Return = $\left[\left(\frac{\mbox{Final Value}}{\mbox{Initial Value}}\right)^{1/n}-1\right]\times 100,$ where $n$ is the number of years, }
    \label{tab:sp500comparison2}
\end{table}

\begin{table}[ht]
    \centering
    \begin{tabular}{c|c|c|c|c|c|c|c|c|c|}
        Year & Method &  \multicolumn{2}{|c|}{Total Return }  &  \multicolumn{2}{|c|}{Annual Return }  & \multicolumn{2}{|c|}{Annual Volatility }  & \multicolumn{2}{|c|}{Sharpe Ratio} \\
        && $n=20$ & $n=50$ & $n=20$& $n=50$ & $n=20$ & $n=30$ & $n=20$ & $n=50$ \\ \hline\hline 
        2013 & PM+MP &33.70&3.84&30.24&6.07&14.58&21.38&2.07&0.28 \\
        & PM+MPNS &26.54&27.09&24.75&24.97&14.88&13.36&1.66&1.87 \\
        & PM+EWP &35.54&40.19&31.77&34.92&15.60&13.99&2.04&2.50 \\\cline{2-10} 
        & MP &\multicolumn{2}{|c|}{-3.38}&\multicolumn{2}{|c|}{-2.57}&\multicolumn{2}{|c|}{13.32}&\multicolumn{2}{|c|}{-0.19}  \\
        & MPNS &\multicolumn{2}{|c|}{26.13}&\multicolumn{2}{|c|}{24.13}&\multicolumn{2}{|c|}{12.71}&\multicolumn{2}{|c|}{1.90} \\
        & EWP &\multicolumn{2}{|c|}{29.84}&\multicolumn{2}{|c|}{26.94}&\multicolumn{2}{|c|}{11.92}&\multicolumn{2}{|c|}{2.26} \\ \hline\hline
2014 & PM+MP &-58.12&-28.15&-38.51&-25.70&98.85&38.64&-0.39& -0.66\\
        & PM+MPNS &11.23&16.23&11.94&16.12&15.78&14.27&0.76&1.13 \\
        & PM+EWP &16.56&15.18&16.88&15.36&17.23&15.20&0.98&1.01 \\\cline{2-10} 
        & MP &\multicolumn{2}{|c|}{-0.40}&\multicolumn{2}{|c|}{0.57}&\multicolumn{2}{|c|}{13.91}&\multicolumn{2}{|c|}{0.04}  \\
        & MPNS &\multicolumn{2}{|c|}{14.84}&\multicolumn{2}{|c|}{14.80}&\multicolumn{2}{|c|}{13.45}&\multicolumn{2}{|c|}{1.10} \\
        & EWP &\multicolumn{2}{|c|}{13.66}&\multicolumn{2}{|c|}{13.56}&\multicolumn{2}{|c|}{11.87}&\multicolumn{2}{|c|}{1.14} \\ \hline \hline
2015 & PM+MP &3.35&-0.18&6.10&11.87&23.56&49.08&0.26&0.24 \\
        & PM+MPNS &6.17&7.57&7.48&8.67&17.12&16.36&0.44&0.53 \\
        & PM+EWP &12.62&2.74&13.36&3.85&16.85&15.05&0.79&0.26 \\\cline{2-10} 
        & MP &\multicolumn{2}{|c|}{-10.18}&\multicolumn{2}{|c|}{-9.32}&\multicolumn{2}{|c|}{17.06}&\multicolumn{2}{|c|}{-0.55}  \\
        & MPNS &\multicolumn{2}{|c|}{5.38}&\multicolumn{2}{|c|}{6.35}&\multicolumn{2}{|c|}{14.76}&\multicolumn{2}{|c|}{0.43} \\
        & EWP &\multicolumn{2}{|c|}{-0.26}&\multicolumn{2}{|c|}{0.91}&\multicolumn{2}{|c|}{15.23}&\multicolumn{2}{|c|}{0.06} \\ \hline \hline
2016 & PM+MP &-1.95&-7.70&0.73&-6.04&23.24&20.03&0.03&-0.30 \\
        & PM+MPNS &-3.57&0.60&-2.51&1.50&15.06&13.46&-0.17&0.11 \\
        & PM+EWP &5.42&9.33&6.62&9.99&16.21&14.35&0.41&0.70 \\\cline{2-10} 
        & MP &\multicolumn{2}{|c|}{13.90}&\multicolumn{2}{|c|}{15.93}&\multicolumn{2}{|c|}{23.95}&\multicolumn{2}{|c|}{0.67}  \\
        & MPNS &\multicolumn{2}{|c|}{3.98}&\multicolumn{2}{|c|}{4.72}&\multicolumn{2}{|c|}{12.64}&\multicolumn{2}{|c|}{0.37} \\
        & EWP &\multicolumn{2}{|c|}{14.92}&\multicolumn{2}{|c|}{15.03}&\multicolumn{2}{|c|}{14.54}&\multicolumn{2}{|c|}{1.03}
        \\ \hline \hline
2017 & PM+MP &5.36&288.53&8.28&228.18&24.52&134.44&0.34&1.70 \\
        & PM+MPNS &7.05&17.54&8.14&16.73&15.95&9.28&0.51&1.80 \\
        & PM+EWP &30.46&25.86&27.59&23.73&12.42&10.32&2.22&2.30 \\\cline{2-10} 
        & MP &\multicolumn{2}{|c|}{49.23}&\multicolumn{2}{|c|}{41.62}&\multicolumn{2}{|c|}{15.73}&\multicolumn{2}{|c|}{2.65}  \\
        & MPNS &\multicolumn{2}{|c|}{17.98}&\multicolumn{2}{|c|}{17.00}&\multicolumn{2}{|c|}{8.16}&\multicolumn{2}{|c|}{2.08} \\
        & EWP &\multicolumn{2}{|c|}{18.73}&\multicolumn{2}{|c|}{17.57}&\multicolumn{2}{|c|}{7.26}&\multicolumn{2}{|c|}{2.42}  \\ \hline \hline
2018 & PM+MP &-23.90&-7.04&-24.79&-5.25&23.30&20.50&-1.06&-0.26 \\
        & PM+MPNS &-17.90&-11.80&-18.05&-11.08&19.11&17.69&-0.94&-0.63 \\
        & PM+EWP &-12.10&-9.39&-10.38&-7.70&22.86&21.13&-0.45&-0.36 \\\cline{2-10} 
        & MP &\multicolumn{2}{|c|}{0.35}&\multicolumn{2}{|c|}{1.81}&\multicolumn{2}{|c|}{17.06}&\multicolumn{2}{|c|}{0.11}  \\
        & MPNS &\multicolumn{2}{|c|}{-7.75}&\multicolumn{2}{|c|}{-6.97}&\multicolumn{2}{|c|}{15.20}&\multicolumn{2}{|c|}{-0.46} \\
        & EWP &\multicolumn{2}{|c|}{-10.03}&\multicolumn{2}{|c|}{-9.38}&\multicolumn{2}{|c|}{15.93}&\multicolumn{2}{|c|}{-0.59} \\ \hline %\hline
    \end{tabular}
   % \caption{SP500 comparison, setting target return as average of the 75‐quartile mean returns } 
    %Total Return $=\frac{\mbox{Final Value} \,-\, \mbox{Initial Value}}{\mbox{Initial Value}}\times 100$, Annual Return = $\left[\left(\frac{\mbox{Final Value}}{\mbox{Initial Value}}\right)^{1/n}-1\right]\times 100,$ where $n$ is the number of years, }
    \label{tab:sp500comparison3}
\end{table}

\begin{table}[ht]
    \centering
    \begin{tabular}{c|c|c|c|c|c|c|c|c|c|}
        Year & Method &  \multicolumn{2}{|c|}{Total Return }  &  \multicolumn{2}{|c|}{Annual Return }  & \multicolumn{2}{|c|}{Annual Volatility }  & \multicolumn{2}{|c|}{Sharpe Ratio} \\
        && $n=20$ & $n=50$ & $n=20$& $n=50$ & $n=20$ & $n=30$ & $n=20$ & $n=50$ \\ \hline\hline
        %\multicolumn{3}{|c|}{Header for all columns} 
2019 & PM+MP &73.80&31.21&64.70&33.11&42.66&34.06&1.52&0.97 \\
        & PM+MPNS &17.16&26.85&16.61&24.40&11.89&10.04&1.40&2.43 \\
        & PM+EWP &18.21&22.65&18.02&21.30&15.60&12.56&1.15&1.70 \\\cline{2-10} 
        & MP &\multicolumn{2}{|c|}{22.74}&\multicolumn{2}{|c|}{23.25}&\multicolumn{2}{|c|}{23.17}&\multicolumn{2}{|c|}{1.00}  \\
        & MPNS &\multicolumn{2}{|c|}{27.24}&\multicolumn{2}{|c|}{24.68}&\multicolumn{2}{|c|}{9.87}&\multicolumn{2}{|c|}{2.50} \\
        & EWP &\multicolumn{2}{|c|}{26.18}&\multicolumn{2}{|c|}{24.16}&\multicolumn{2}{|c|}{12.63}&\multicolumn{2}{|c|}{1.91} \\ \hline\hline
2020 & PM+MP &-81.08&-26.83&-94.94&14.99&123.30&96.72&-0.77&0.15 \\
        & PM+MPNS &4.90&4.60&12.18&11.10&38.24&36.12&0.32&0.31 \\
        & PM+EWP &9.00&11.49&16.01&17.27&38.14&35.54&0.42& 0.49\\\cline{2-10} 
        & MP &\multicolumn{2}{|c|}{-5.76}&\multicolumn{2}{|c|}{3.28}&\multicolumn{2}{|c|}{42.30}&\multicolumn{2}{|c|}{0.08}  \\
        & MPNS &\multicolumn{2}{|c|}{2.80}&\multicolumn{2}{|c|}{9.32}&\multicolumn{2}{|c|}{35.98}&\multicolumn{2}{|c|}{0.26} \\
        & EWP &\multicolumn{2}{|c|}{-1.60}&\multicolumn{2}{|c|}{5.61}&\multicolumn{2}{|c|}{37.69}&\multicolumn{2}{|c|}{0.15} \\ \hline \hline
2021 & PM+MP &-35.45&8.14&80.17&11.58&157.42&27.30&0.51&0.42 \\
        & PM+MPNS &19.54&11.27&19.81&11.66&19.38&13.68&1.02&0.85 \\
        & PM+EWP &23.16&24.38&22.53&23.04&17.85&14.99&1.26&1.54 \\\cline{2-10} 
        & MP &\multicolumn{2}{|c|}{-1.97}&\multicolumn{2}{|c|}{0.43}&\multicolumn{2}{|c|}{22.04}&\multicolumn{2}{|c|}{0.02}  \\
        & MPNS &\multicolumn{2}{|c|}{13.67}&\multicolumn{2}{|c|}{13.60}&\multicolumn{2}{|c|}{12.07}&\multicolumn{2}{|c|}{1.13} \\
        & EWP &\multicolumn{2}{|c|}{24.31}&\multicolumn{2}{|c|}{22.74}&\multicolumn{2}{|c|}{13.30}&\multicolumn{2}{|c|}{1.71} \\ \hline \hline
2022 & PM+MP &-100.00&-99.30&353.52&142.60&843.63&354.21&0.42&0.40 \\
        & PM+MPNS &-24.14&-22.28&-25.92&-23.96&26.22&23.76&-0.99&-1.01 \\
        & PM+EWP &-23.10&-20.00&-23.78&-20.46&28.70&25.51&-0.83&-0.80 \\\cline{2-10} 
        & MP &\multicolumn{2}{|c|}{-8.19}&\multicolumn{2}{|c|}{-5.72}&\multicolumn{2}{|c|}{25.94}&\multicolumn{2}{|c|}{-0.22}  \\
        & MPNS &\multicolumn{2}{|c|}{-18.23}&\multicolumn{2}{|c|}{-19.11}&\multicolumn{2}{|c|}{21.39}&\multicolumn{2}{|c|}{-0.89} \\
        & EWP &\multicolumn{2}{|c|}{-17.62}&\multicolumn{2}{|c|}{-17.91}&\multicolumn{2}{|c|}{23.19}&\multicolumn{2}{|c|}{-0.77} \\ \hline \hline
    \end{tabular}
   % \caption{SP500 comparison, setting target return as average of the 75‐quartile mean returns } 
    %Total Return $=\frac{\mbox{Final Value} \,-\, \mbox{Initial Value}}{\mbox{Initial Value}}\times 100$, Annual Return = $\left[\left(\frac{\mbox{Final Value}}{\mbox{Initial Value}}\right)^{1/n}-1\right]\times 100,$ where $n$ is the number of years, }
    \label{tab:sp500comparison4}
\end{table}

We also observe an interesting phenomena while solving OMV1 for the Market Champions dataset. From Table \ref{tab:comparison}, note that the values of all the statistics obtained for Markowitz's no short selling method (MPNS) match with the corresponding statistics obtained by employing our proposed method combining hedge scores of the assets and the MPNS. This happened due to the fact that the naive MPNS for all the assets construct the same portfolio as $\mathcal{S}_5.$ Thus we see a deep connection of hedge scores with the solution of MPNS that makes our proposed dimension reduction technique using hedge statistics important and opens an avenue for future research. \\

\clearpage
\noindent\textbf{Conclusion:} In this paper, we propose a method for dimensionality reduction of Markowitz’s mean–variance portfolio optimization problem by modeling the local dynamics of asset returns through a signed graph framework.  Specifically, we define the hedge-score of an asset in terms of the negative degree of the corresponding vertex in the graph representation of the financial market. To evaluate the effectiveness of this approach, we conduct backtesting on two datasets and benchmark the performance of the proposed method on the reduced asset universe against that of Markowitz’s optimization (with and without short selling) as well as the equally weighted portfolio on the full universe.

Our empirical analysis shows that the proposed method outperforms the standard approaches on several occasions, thereby demonstrating its potential efficiency. However, in other cases it fails to achieve comparable performance. Such variability may arise from factors including the choice of $K$, the number of potential hedge-protected assets to be selected. In future work, we intend to explore the integration of higher-order motifs in the signed graph framework as a means of further enhancing dimensionality reduction.   \\

\noindent{\bf Acknowledgment.} The author thanks Sarvagya Upadhyay and Hannes Leipold for inspiring discussions and valuable suggestions.

%\begin{thebibliography}{6}
%

%\bibitem {bib1}
%Smith, T.F., Waterman, M.S.: Identification of common molecular subsequences.
%J. Mol. Biol. 147, 195?197 (1981). \url{doi:10.1016/0022-2836(81)90087-5}

%\bibitem {may:ehr:stein}
%May, P., Ehrlich, H.-C., Steinke, T.: ZIB structure prediction pipeline:
%composing a complex biological workflow through web services. In: Nagel, W.E., Walter, W.V., Lehner, W. (eds.) Euro-Par 2006. LNCS, vol. 4128, pp. 1148?1158. Springer, Heidelberg (2006).
%\url{doi:10.1007/11823285_121}

%\bibitem {fost:kes}
%Foster, I., Kesselman, C.: The Grid: Blueprint for a New Computing Infrastructure.
%Morgan Kaufmann, San Francisco (1999)

%\bibitem {czaj:fitz} Czajkowski, K., Fitzgerald, S., Foster, I., Kesselman, C.: Grid information servicesfor distributed resource sharing. In: 10th IEEE International Symposiumon High Performance Distributed Computing, pp. 181?184. IEEE Press, New York (2001).\url{doi: 10.1109/HPDC.2001.945188}

%\bibitem {fo:kes:nic:tue}
%Foster, I., Kesselman, C., Nick, J., Tuecke, S.: The physiology of the grid: an open grid services architecture for distributed systems integration. Technical report, Global GridForum (2002)

%\bibitem {onlyurl}
%National Center for Biotechnology Information. \url{http://www.ncbi.nlm.nih.gov}

%\end{thebibliography}

% ---- OR ------- Uncomment this section to use bibtex
% \bibliographystyle{spmpsci} % We choose the "plain" reference style
% \bibliography{refs} % Entries are in the refs.bib file

\begin{thebibliography}{6}

\bibitem{acemoglu2015systemic}
Daron Acemoglu, Asuman Ozdaglar, and Alireza Tahbaz-Salehi.
\newblock Systemic risk and stability in financial networks.
\newblock  American Economic Review, 105(2):564--608, 2015.

\bibitem{aref2019balance}
Samin Aref and Mark~C Wilson.
\newblock Balance and frustration in signed networks.
\newblock Journal of Complex Networks, 7(2):163--189, 2019.

\bibitem{barigozzi2019nets}
Matteo Barigozzi and Christian Brownlees.
\newblock Nets: Network estimation for time series.
\newblock  Journal of Applied Econometrics, 34(3):347--364, 2019.

\bibitem{bartesaghi2025global}
Paolo Bartesaghi, Fernando Diaz-Diaz, Rosanna Grassi, and Pierpaolo Uberti.
\newblock Global balance and systemic risk in financial correlation networks.
\newblock  Physica A: Statistical Mechanics and its Applications, page 130698, 2025.

\bibitem{baur2010gold}
Dirk~G Baur and Brian~M Lucey.
\newblock Is gold a hedge or a safe haven? an analysis of stocks, bonds and gold.
\newblock Financial review, 45(2):217--229, 2010.

\bibitem{bollerslev1986generalized}
Tim Bollerslev.
\newblock Generalized autoregressive conditional heteroskedasticity.
\newblock Journal of econometrics, 31(3):307--327, 1986.

\bibitem{boyd2024markowitz}
Stephen Boyd, Kasper Johansson, Ronald Kahn, Philipp Schiele, and Thomas Schmelzer.
\newblock Markowitz portfolio construction at seventy.
\newblock arXiv preprint arXiv:2401.05080, 2024.

\bibitem{buonaiuto2023best}
Giuseppe Buonaiuto, Francesco Gargiulo, Giuseppe De~Pietro, Massimo Esposito, and Marco Pota.
\newblock Best practices for portfolio optimization by quantum computing, experimented on real quantum devices.
\newblock  Scientific Reports, 13(1):19434, 2023.

\bibitem{cartwright1956structural}
Dorwin Cartwright and Frank Harary.
\newblock Structural balance: a generalization of heider's theory.
\newblock Psychological review, 63(5):277, 1956.

\bibitem{chen2020correlation}
Lin Chen, Qian Han, Zhilin Qiao, and H. Eugene Stanley. 
\newblock Correlation analysis and systemic risk measurement of regional, financial and global stock indices.
\newblock Physica A: Statistical Mechanics and its Applications 542:122653, 2020.

\bibitem{chi2010network}
K~Tse Chi, Jing Liu, and Francis~CM Lau.
\newblock A network perspective of the stock market.
\newblock { Journal of Empirical Finance}, 17(4):659--667, 2010.

\bibitem{chung2022effects}
Munki Chung, Yongjae Lee, Jang~Ho Kim, Woo~Chang Kim, and Frank~J Fabozzi.
\newblock The effects of errors in means, variances, and correlations on the mean-variance framework.
\newblock { Quantitative Finance}, 22(10):1893--1903, 2022.

\bibitem{demiguel2009optimal}
Victor DeMiguel, Lorenzo Garlappi, and Raman Uppal. \newblock Optimal versus naive diversification: How inefficient is the 1/N portfolio strategy?. 
\newblock {The review of Financial studies} 22(5):1915-1953, 2009.

\bibitem{diaz2025mathematical}
Fernando Diaz-Diaz.
\newblock { Mathematical analysis of signed networks: structure and dynamics}.
\newblock PhD thesis, Institute of Cross-Disciplinary Physics and Complex Systems, IFISC, 2025.

\bibitem{ehsani2020structure}
Maryam Ehsani.
\newblock The structure of stock markets as signed networks.
\newblock { Journal of Industrial and Systems Engineering}, 13(1):136--146, 2020.

%\bibitem{elliott2014financial}
%Matthew Elliott, Benjamin Golub, and Matthew~O Jackson.
%\newblock Financial networks and contagion.
%\newblock { American Economic Review}, 104(10):3115--3153, 2014.

\bibitem{engle1982autoregressive}
Robert~F Engle.
\newblock Autoregressive conditional heteroscedasticity with estimates of the variance of united kingdom inflation.
\newblock { Econometrica: Journal of the econometric society}, pages 987--1007, 1982.

\bibitem{figueiredo2014maximum}
Rosa Figueiredo and Yuri Frota.
\newblock The maximum balanced subgraph of a signed graph: Applications and solution approaches.
\newblock { European Journal of Operational Research}, 236(2):473--487, 2014.

\bibitem{goldberg2022dispersion}
Lisa~R Goldberg, Alex Papanicolaou, and Alex Shkolnik.
\newblock The dispersion bias.
\newblock { SIAM Journal on Financial Mathematics}, 13(2):521--550, 2022.

\bibitem{gregnanin2024signature}
Marco Gregnanin, Yanyi Zhang, Johannes De Smedt, Giorgio Gnecco, and Maurizio Parton. 
\newblock Signature-based portfolio allocation: a network approach. 
\newblock {Applied Network Science} 9, no. 1: 54,  2024.

\bibitem{guhr2003new}
Thomas Guhr and Bernd K{\"a}lber.
\newblock A new method to estimate the noise in financial correlation matrices.
\newblock { Journal of Physics A: Mathematical and General}, 36(12):3009, 2003.

\bibitem{harary1953notion}
Frank Harary.
\newblock On the notion of balance of a signed graph.
\newblock { Michigan Mathematical Journal}, 2(2):143--146, 1953.

\bibitem{harary2002signed}
Frank Harary, Meng-Hiot Lim, and Donald~C Wunsch.
\newblock Signed graphs for portfolio analysis in risk management.
\newblock { IMA Journal of management mathematics}, 13(3):201--210, 2002.

\bibitem{hautsch2015financial}
Nikolaus Hautsch, Julia Schaumburg, and Melanie Schienle.
\newblock Financial network systemic risk contributions.
\newblock { Review of Finance}, 19(2):685--738, 2015.

\bibitem{hegade2022portfolio}
Narendra~N Hegade, Pranav Chandarana, Koushik Paul, Xi~Chen, Francisco Albarr{\'a}n-Arriagada, and E~Solano.
\newblock Portfolio optimization with digitized counterdiabatic quantum algorithms.
\newblock { Physical Review Research}, 4(4):043204, 2022.

\bibitem{huffner2010separator}
Falk H{\"u}ffner, Nadja Betzler, and Rolf Niedermeier.
\newblock Separator-based data reduction for signed graph balancing.
\newblock { Journal of combinatorial optimization}, 20(4):335--360, 2010.

\bibitem{jackson2014networks}
Matthew~O Jackson.
\newblock Networks in the understanding of economic behaviors.
\newblock { Journal of economic perspectives}, 28(4):3--22, 2014.

\bibitem{kendall1938new}
Maurice~G Kendall.
\newblock A new measure of rank correlation.
\newblock { Biometrika}, 30(1-2):81--93, 1938.

\bibitem{kenett2015network}
Dror~Y Kenett and Shlomo Havlin.
\newblock Network science: a useful tool in economics and finance.
\newblock { Mind \& Society}, 14:155--167, 2015.

\bibitem{kondor2007noise}
Imre Kondor, Szil{\'a}rd Pafka, and G{\'a}bor Nagy.
\newblock Noise sensitivity of portfolio selection under various risk measures.
\newblock { Journal of Banking \& Finance}, 31(5):1545--1573, 2007.

\bibitem{kordonowy2025lie}
Steven Kordonowy and Hannes Leipold.
\newblock The lie algebra of xy-mixer topologies and warm starting qaoa for constrained optimization.
\newblock { arXiv preprint arXiv:2505.18396}, 2025.

\bibitem{lai2022survey}
Zhao-Rong Lai and Haisheng Yang.
\newblock A survey on gaps between mean-variance approach and exponential growth rate approach for portfolio optimization.
\newblock { ACM Computing Surveys (CSUR)}, 55(2):1--36, 2022.

\bibitem{laloux1999noise}
Laurent Laloux, Pierre Cizeau, Jean-Philippe Bouchaud, and Marc Potters.
\newblock Noise dressing of financial correlation matrices.
\newblock { Physical review letters}, 83(7):1467, 1999.

\bibitem{leipold2024train}
Hannes Leipold and Sarvagya Upadhyay.
\newblock Train-and-scaling the quantum alternating operator ansatz to solve portfolio diversification.
\newblock In { 2024 IEEE International Conference on Quantum Computing and Engineering (QCE)}, volume~2, pages 132--137. IEEE, 2024.

\bibitem{mantegna1999hierarchical}
Rosario~N Mantegna.
\newblock Hierarchical structure in financial markets.
\newblock { The European Physical Journal B-Condensed Matter and Complex Systems}, 11:193--197, 1999.

\bibitem{mantegna1999introduction}
Rosario N. Mantegna and H. Eugene Stanley. \newblock Introduction to econophysics: correlations and complexity in finance. \newblock Cambridge university press, 1999.

\bibitem{markowitz1952portfolio}
Harry~M Markowitz.
\newblock Portfolio selection, the journal of finance. 7 (1).
\newblock { N}, 1:71--91, 1952.

\bibitem{masuda2025introduction}
 Naoki Masuda, Zachary M. Boyd, Diego Garlaschelli, and Peter J. Mucha. 
\newblock Introduction to correlation networks: Interdisciplinary approaches beyond thresholding.
\newblock {Physics Reports} 1136: 1-39, 2025.

\bibitem{mattera2023shrinkage}
Giulio Mattera and Raffaele Mattera.
\newblock Shrinkage estimation with reinforcement learning of large variance matrices for portfolio selection.
\newblock { Intelligent Systems with Applications}, 17:200181, 2023.

%\bibitem{nagurney2021networks}
%Anna Nagurney.
%\newblock Networks in economics and finance in networks and beyond: A half century retrospective.
%\newblock { Networks}, 77(1):50--65, 2021.

\bibitem{pafka2003noisy}
Szil{\'a}rd Pafka and Imre Kondor.
\newblock Noisy covariance matrices and portfolio optimization ii.
\newblock { Physica A: Statistical Mechanics and its Applications}, 319:487--494, 2003.

%\bibitem{pafka2004estimated}
%Szilard Pafka and Imre Kondor.
%\newblock Estimated correlation matrices and portfolio optimization.
%\newblock { Physica A: statistical mechanics and its applications}, 343:623--634, 2004.

\bibitem{peralta2016network}
Gustavo Peralta and Abalfazl Zareei.
\newblock A network approach to portfolio selection.
\newblock { Journal of Empirical Finance}, 38:157--180, 2016.

\bibitem{singh2017measuring}
Ranveer Singh and Bibhas Adhikari.
\newblock Measuring the balance of signed networks and its application to sign prediction.
\newblock { Journal of Statistical Mechanics: Theory and Experiment}, 2017(6):063302, 2017.

\bibitem{soloviev2025scaling}
Soloviev, Vicente P., Antonio Márquez Romero, Josh Kirsopp, and Michal Krompiec. 
\newblock Scaling Portfolio Diversification with Quantum Circuit Cutting Techniques.
\newblock {arXiv preprint} arXiv:2506.08947, 2025.

\bibitem{zaslavsky2012mathematical}
Thomas Zaslavsky.
\newblock A mathematical bibliography of signed and gain graphs and allied areas.
\newblock { The Electronic Journal of Combinatorics}, pages DS8--Dec, 2012.

\bibitem{vasanthi2015applications}
B. Vasanthi, S. Arumugam, Atulya K. Nagar, and Sovan Mitra. 
\newblock Applications of signed graphs to portfolio turnover analysis. 
\newblock {Procedia-Social and Behavioral Sciences} 211: 1203-1209, 2015.

\end{thebibliography}
\end{document}